\documentclass[final,12pt]{clear2026} 
\usepackage{caption}
\usepackage{graphicx} 
\usepackage{psfrag,epsf}
\usepackage{enumerate}
\usepackage{bbm}
\usepackage{hyperref}
\usepackage{cleveref}
\usepackage{bm}
\usepackage{algorithm}
\usepackage{algpseudocode}
\usepackage{booktabs}
\usepackage{multicol}
\usepackage{multirow}
\usepackage{xcolor}
\usepackage{url} 
\usepackage{wrapfig}

\DeclareMathOperator*{\E}{\mathbb{E}}
\DeclareMathOperator*{\R}{\mathbb{R}}
\renewcommand{\P}{\mathbb{P}}

\newtheorem{assumption}[theorem]{Assumption}
\newcommand{\qed}{\hfill$\blacksquare$}

\title[Sharp Generalization Bounds under Outcome Shift]{Sharp Bounds for Treatment Effect Generalization under Outcome Distribution Shift}
\usepackage{times}

\clearauthor{%
 \Name{Amir Asiaee} \Email{amir.asiaeetaheri@vumc.org}\\
 \Name{Samhita Pal} \Email{samhita.pal@vumc.org}\\
 \Name{Cole Beck} \Email{cole.beck@vumc.org}\\
 \addr Department of Biostatistics, Vanderbilt University Medical Center, Nashville, TN, USA
 \AND
 \Name{Jared D. Huling} \Email{huling@umn.edu}\\
 \addr Division of Biostatistics and Health Data Science, University of Minnesota, Minneapolis, MN, USA%
}

\begin{document}

\maketitle

\noindent\textbf{Note:} Published in \emph{Proceedings of the 5th Conference on Causal Learning and Reasoning (CLeaR 2026)}, PMLR.\medskip

\begin{abstract}
Generalizing treatment effects from a randomized trial to a target population requires
the assumption that potential outcome distributions are invariant across populations
after conditioning on observed covariates. This assumption fails when unmeasured effect
modifiers are distributed differently between trial participants and the target
population. We develop a sensitivity analysis framework that bounds how much conclusions
can change when this transportability assumption is violated. Our approach constrains
the likelihood ratio between target and trial outcome densities by a scalar parameter
$\Lambda \geq 1$, with $\Lambda = 1$ recovering standard transportability. For each
$\Lambda$, we derive sharp bounds on the target average treatment effect---the tightest
interval guaranteed to contain the true effect under all data-generating processes
compatible with the observed data and the sensitivity model. We show that the optimal
likelihood ratios have a simple threshold structure, leading to a closed-form greedy
algorithm that requires only sorting trial outcomes and redistributing probability mass.
The resulting estimator runs in $O(n \log n)$ time and is consistent under standard
regularity conditions. Simulations demonstrate that our bounds achieve nominal coverage
when the true outcome shift falls within the specified $\Lambda$, provide substantially
tighter intervals than worst-case bounds, and remain informative across a range of
realistic violations of transportability.
\end{abstract}


\section{Introduction}
\label{sec:intro}

Randomized controlled trials (RCTs) are widely regarded as the gold standard for estimating causal effects because randomization yields strong internal validity. Yet trial participants often differ systematically from the real-world patients or communities to which results are ultimately applied, raising concerns about \emph{external validity}. A large methodological literature studies how to \emph{generalize} or \emph{transport} treatment effects from a trial to a target population by combining trial data with covariate information from the target population, typically via outcome-model-based standardization (g-formula),
inverse-odds-of-participation weighting, or augmented/doubly robust
estimators that combine both \citep{cole2010generalizing,stuart2011use,
buchanan2018generalizing,dahabreh2019extending}. Closely related identification questions appear in the causal graphical literature under \emph{transportability} and \emph{data fusion} \citep{pearl2011transportability,bareinboim2016causal}. Recent reviews summarize the resulting toolkit and its applied use across disciplines \citep{degtiar2023review,ling2023overview,colnet2024causal}.

We study a canonical trial generalization design with two data sources: a randomized trial and a sample intended to represent the target population. For individuals enrolled in the trial, we observe baseline covariates, randomized treatment assignment, and the outcome. For individuals in the target population sample, we observe the same baseline covariates but not treatment assignments or outcomes. Our inferential goal is the \emph{target average treatment effect} (TATE), i.e., the average causal effect of the trial intervention in the target population. We defer formal notation and the precise definition of the estimand to Section~\ref{sec:setup}.

Standard generalization estimators rely on assumptions that link the trial to the target population. Informally, a common sufficient condition is that, after adjusting for the observed baseline covariates, trial participation is independent of the potential outcomes, so that remaining differences between the trial and the target population are fully explained by their covariate distributions. In addition, identification requires adequate overlap in covariate support between the two populations and the usual causal assumptions for a randomized trial (e.g., consistency and randomization). Under these conditions, the TATE is identified either by reweighting trial participants to represent the target population or by outcome-model-based standardization to the target covariate distribution \citep{cole2010generalizing,stuart2011use,buchanan2018generalizing,dahabreh2021study}. Terminology and precise assumptions vary across communities; see \citet{dahabreh2019extendingb} and the synthesis in \citet{degtiar2023review}.

In practice, however, conditional exchangeability across participation is not testable from the observed data and can fail even when rich covariates are available. Multiple mechanisms can cause the conditional outcome distribution to differ between trial and target populations. The most commonly discussed is \emph{unmeasured effect modification}: trial participation may depend on variables that also modify treatment effects, and whose distributions differ across populations. But outcome shift can also arise from \emph{treatment version variability} (the intervention implemented in routine care may differ from the trial protocol \citep{vanderweele2009concerning}) or \emph{network interference} (outcomes in the target may depend on community-level treatment patterns absent in the trial \citep{hudgens2008toward}). All three mechanisms produce the same observable consequence---a shift in conditional outcome distributions between populations---and motivate \emph{sensitivity analyses} that quantify how conclusions change under controlled departures from outcome exchangeability. Several recent proposals develop sensitivity analyses for omitted moderators---including settings where a moderator is observed in the trial but unavailable in the target dataset  \citep{nguyen2018sensitivity,nie2021covariate,dahabreh2023sensitivity,huang2024sensitivity}. These approaches work by positing a sensitivity model bounding how unobserved moderators relate to selection and/or treatment effect heterogeneity.

We develop a complementary \emph{distributional} sensitivity analysis for trial generalization that directly constrains how much the \emph{conditional outcome distribution} may differ between the trial and target populations after adjusting for observed covariates and treatment arm. Specifically, we introduce an \emph{outcome-shift} model indexed by a single sensitivity parameter $\Lambda \ge 1$. The model restricts the target conditional outcome distribution to lie within a pointwise likelihood-ratio envelope around the trial distribution. This allows flexible, outcome-dependent changes (including shifts concentrated in the tails), while ruling out arbitrarily sharp local spikes that would make sensitivity intervals uninformative. In particular, this is not ``uniform density inflation/deflation'': the distortion can vary with the outcome value within the envelope. The restriction also implies a corresponding bound on how much conditional means can shift, but not vice versa; we formalize these relationships after introducing notation.
At its baseline value $\Lambda = 1$, the model reduces to the usual outcome-invariance condition underpinning standard generalization methods; as $\Lambda$ increases, it allows progressively larger departures in a controlled way.

Compared to sensitivity models that bound the density ratio of an unmeasured effect modifier across populations \citep{nie2021covariate}, our restriction is placed directly on the induced change in the observed outcome distribution within treatment arms. The model is natural in multi-site clinical trials (where site-level differences in care patterns may alter outcome distributions), in policy generalization (where local context may moderate treatment effects), and in health equity applications (where underrepresented populations may have different outcome distributions due to unmeasured social determinants).
This perspective is closely related in spirit to likelihood-ratio/odds-ratio sensitivity models for unmeasured confounding in observational studies \citep{rosenbaum2002obs,tan2006marginal,zhao2019sensitivity,dorn2023sharp} and to recent work on sharp bounds under generalized causal sensitivity models \citep{franks2020flexible,frauen2023sharp}, but our target is \emph{external validity}: uncertainty enters through possible differences in target-versus-trial outcome distributions rather than by unmeasured confounding of treatment assignment.

Our main contributions are as follows. We derive sharp identification bounds for the target average treatment effect under distributional (pointwise density-ratio) outcome shift between trial and target populations, and we show these bounds can be computed by a closed-form $O(n\log n)$ greedy algorithm. While the mathematical techniques we employ---threshold arguments, linear programming, greedy algorithms---are related to tools used in the marginal sensitivity literature for unmeasured confounding \citep{tan2006marginal,dorn2023sharp}, the problem formulation is different: combining RCT randomization, covariate transport weighting, and distributional outcome shift into a single framework targets external validity rather than internal validity. Concretely:
\begin{itemize}
  \item \textbf{Sharp population-level bounds.} We give a convex characterization of
  the identified set for target mean potential outcomes and the target ATE under the
  outcome-shift sensitivity model. We show that the extremal outcome distributions
  have a simple threshold structure: the optimal likelihood ratio takes only the
  boundary values $\Lambda^{-1}$ and $\Lambda$, switching at a single outcome
  threshold. This yields closed-form expressions for the sharp bounds and clarifies
  how interval width depends on $\Lambda$ and the variability of trial outcomes.

  \item \textbf{A simple estimation algorithm.} At the sample level, we derive a
  discrete optimization formulation for the worst-case target means and show that the
  corresponding linear programs admit a greedy closed-form solution. The algorithm
  requires only sorting trial outcomes, computing generalization weights, and
  redistributing probability mass toward extreme outcomes subject to the likelihood
  ratio constraint. This yields $O(n \log n)$ estimators for the bounds, with
  consistency following from standard empirical process arguments.

  \item \textbf{Interpretable robustness summary.} We introduce a \emph{tipping-point} sensitivity level $\Lambda^\star$: the smallest outcome-shift magnitude at which the identified set first includes the null. This provides a robustness metric analogous to Rosenbaum's $\Gamma$ for observational studies \citep{rosenbaum2005sensitivity}: reporting $\Lambda^\star$ answers ``how large would the outcome distribution shift have to be to overturn the treatment effect conclusion?'' without requiring the analyst to commit to a specific $\Lambda$. We distinguish $\Lambda^\star$ (a population identification quantity) from the empirical ``breakeven'' $\Lambda_{\mathrm{out}}$ reported in our simulations (the smallest $\Lambda$ achieving $\ge 95\%$ coverage), which additionally incorporates finite-sample estimation error.
\end{itemize}

\section{Related work}
\label{sec:related}

\paragraph{Generalizability and transportability of treatment effects.}
Methods for extending causal conclusions beyond the experimental sample are often framed as generalizability or transportability problems. In the potential outcomes literature, identification and estimation typically proceed by modeling or reweighting the trial sample to match the target covariate distribution using a sampling score (trial participation propensity) \citep{cole2010generalizing,stuart2011use,buchanan2018generalizing}, and by clarifying how available data sources (nested vs.\ non-nested designs) affect identification and estimation \citep{dahabreh2019extending,dahabreh2021study}. In the structural causal model literature, transportability is studied via selection diagrams and data fusion, which characterize when and how causal relations can be moved across environments \citep{pearl2011transportability,bareinboim2016causal}. Comprehensive overviews and applied perspectives appear in \citet{degtiar2023review}, \citet{ling2023overview}, and \citet{colnet2024causal}.
A related line of work uses observational data to improve trial-based treatment effect estimation under distribution shift between trial and external populations \citep{asiaee2023roscar,asiaee2025calm,pal2025mroscar,karlsson2025robust}.

\paragraph{Sensitivity analysis for external validity violations.}
All generalization/transport estimators rely on assumptions that connect trial and target populations, most notably conditions ensuring that the set of observed covariates suffices to adjust for selection into the trial. Because this is inherently uncertain in applications, sensitivity analyses have been proposed to assess robustness to omitted moderators or departures from conditional exchangeability across participation. \cite{nguyen2018sensitivity} develop sensitivity analyses when an effect modifier is observed in the trial but not in the target population.
\cite{nie2021covariate} develop a covariate-balancing sensitivity analysis for extrapolating randomized trials across locations, using a bounded conditional density-ratio model for the unmeasured effect modifier across populations to obtain optimization-based identification intervals tightened via covariate moment balancing.
 \cite{dahabreh2023sensitivity} propose bias-function-based sensitivity analyses that directly parameterize violations of generalizability/transportability assumptions in terms of deviations in target counterfactual means. \cite{huang2024sensitivity} proposes a sensitivity analysis for weighted generalization estimators in which the key sensitivity parameters can be expressed as a correlation and an explained-variation measure; these are standardized and scale-invariant, and the paper develops benchmarking tools for plausible parameter values. In a complementary direction, \citet{asiaee2026ovb} develop an omitted-variable-bias decomposition for external validity that yields closed-form sensitivity bounds parameterized by partial $R^2$ values.
Our approach instead bounds the discrepancy between trial and target \emph{conditional outcome distributions} within treatment arms, yielding worst-case identification intervals indexed by a sensitivity parameter $\Lambda$.

\paragraph{Partial identification.}
Our framework is an instance of partial identification in the tradition of \citet{manski1990nonparametric,manski2003partial}. Manski's program asks what can be learned about a causal estimand when the data and maintained assumptions do not pin it down to a single value; the answer is an \emph{identified set} whose width reflects the strength of the maintained assumptions. In our setting, the target ATE is generally only partially identified, and the sensitivity parameter $\Lambda$ controls assumption strength: $\Lambda = 1$ yields point identification (standard transportability), $\Lambda \to \infty$ recovers worst-case bounds from outcome boundedness alone, and intermediate $\Lambda$ traces out an \emph{identification frontier}---the trade-off between assumption strength and inferential precision.

\paragraph{Connection to confounding sensitivity analysis and sharp bounds.}
Sensitivity analysis for unmeasured confounding in observational studies has a long history \citep{rosenbaum2002obs}. A prominent modern formalization is the marginal sensitivity model, which bounds how much the true treatment assignment odds can differ from a nominal propensity score model \citep{tan2006marginal}. Building on such models, Zhao et al.\ \citep{zhao2019sensitivity} and Dorn and Guo \citep{dorn2023sharp} develop sharp bounds and inference procedures for inverse propensity weighting estimators. Other recent work proposes flexible or generalized sensitivity model frameworks and derives sharp bounds under broad classes of sensitivity restrictions \citep{franks2020flexible,frauen2023sharp}. We adapt sharp-bound techniques from this marginal sensitivity literature to the external validity setting. While the mathematical tools are related, the problem formulation is new: combining RCT randomization, covariate transport weighting, and distributional outcome shift into a single framework for the target ATE requires new identification arguments specific to the generalizability setting.

%

\vspace{-12pt}
\section{Problem Formulation and Notation}
\label{sec:setup}

We adopt the potential outcomes framework for multi-population causal inference,
following \citet{dahabreh2019extending} with notation adapted for our sensitivity analysis.
Let $S \in \{r, o\}$ index the \emph{source} randomized trial ($S = r$) and the
\emph{target} population ($S = o$). For unit $i$ in population $s$, we observe a
covariate vector $X_i^s \in \mathcal{X} \subset \mathbb{R}^p$, a binary treatment
$A_i^s \in \{-1, +1\}$, and an outcome $Y_i^s \in \mathcal{Y} \subset \mathbb{R}$.
Potential outcomes are denoted $(Y_i(+1), Y_i(-1))$, with the usual consistency
assumption $Y_i = Y_i(A_i)$.

Throughout, we use the study indicator $s \in \{r, o\}$ as a superscript to denote
quantities specific to each population. When we write $P^s$, $\mathbb{E}^s[\cdot]$,
or $\mu_a^s$, this should be understood as the distribution, expectation, or
parameter \emph{conditional on membership in study $s$}---that is, we are
characterizing the data-generating process within each population separately.

For treatment level $a \in \{-1, +1\}$, we define the conditional and marginal
mean potential outcomes in population $s$ as
$\mu_a^s(x) := \mathbb{E}^s[Y(a) \mid X = x]$ and $\mu_a^s := \mathbb{E}^s[Y(a)]$.
The average treatment effect (ATE) in population $s$ is
$\tau^s := \mu_{+1}^s - \mu_{-1}^s = \mathbb{E}^s[Y(+1) - Y(-1)]$.
Our goal is to \emph{generalize} the trial-based treatment effect to the target
population---specifically, to learn about $\tau^o = \mathbb{E}^o[Y(+1) - Y(-1)]$---using
data from the trial together with covariate information from the target population.

We observe an i.i.d.\ sample $\mathcal{D}^r := \{(X_i^r, A_i^r, Y_i^r)\}_{i=1}^{n^r}$
from the trial, where treatment is randomized, and a separate i.i.d.\ sample of
covariates $\mathcal{D}^o := \{X_j^o\}_{j=1}^{n^o}$ from the target population.
Crucially, we do not observe outcomes or treatment assignments in the target sample,
so the conditional outcome distribution in the target, $P^o(Y \mid A, X)$, is not
directly identified from the data.

\subsection{Identification under standard transportability.}
To isolate the role of outcome shift, we assume the usual conditions for internal
validity of the trial.

\begin{assumption}[Trial internal validity]
\label{assump:internal-validity}
For the trial population $S = r$:
\begin{enumerate}[(i)]
  \item Consistency: $Y = Y(A)$ almost surely.
  \item Randomization: $(Y(+1), Y(-1)) \perp A \mid X, S = r$.
  \item Positivity: there exists $\eta > 0$ such that
        $\mathbb{P}^r(A = a \mid X = x) \geq \eta$ for all $a \in \{-1, +1\}$
        and almost all $x$.
\end{enumerate}
\end{assumption}

Under Assumption~\ref{assump:internal-validity}, the conditional mean potential
outcome is identified from trial data as $\mu_a^r(x) = \mathbb{E}^r[Y \mid A = a, X = x]$.

\paragraph{No outcome shift.}
The standard approach to generalization further assumes \emph{outcome transportability}:
the conditional distribution of potential outcomes is identical across populations,
\begin{equation}
  P^r(Y(a) \mid X = x) = P^o(Y(a) \mid X = x) \quad \text{for all } x \text{ and } a,
  \label{eq:transportability}
\end{equation}
sometimes written compactly as $Y(a) \perp S \mid X$ \citep{pearl2011transportability,bareinboim2016causal}.
Under this assumption, combining trial outcomes with the target covariate distribution
identifies the target effect. Let $w(x) := dP^{o} / dP^{r}(x)$ denote the density
ratio between the target and trial covariate distributions. Then
\begin{equation}
  \tau^o = \mathbb{E}^o\bigl[\mu_{+1}^r(X) - \mu_{-1}^r(X)\bigr]
         = \mathbb{E}^r\bigl[w(X) \cdot (\mu_{+1}^r(X) - \mu_{-1}^r(X))\bigr].
  \label{eq:tau-transport}
\end{equation}
This identity suggests two estimation strategies \citep{stuart2011use,dahabreh2019extending}.
The \emph{outcome modeling} approach fits regression models $\hat{\mu}_a^r(x)$ on
trial data and averages over the target covariate sample:
$\hat{\tau}^o_{\text{om}} = (n^o)^{-1} \sum_{j=1}^{n^o} [\hat{\mu}_{+1}^r(X_j^o) - \hat{\mu}_{-1}^r(X_j^o)]$.
The \emph{inverse probability weighting} approach estimates the density ratio
$\hat{w}(x)$ typically via logistic regression of the study indicator $S$ on $X$
using pooled data
and computes arm-specific weighted means,
\(
  \hat{\mu}_{a,\text{ipw}}^o
  :=
  \frac{\sum_{i=1}^{n^r} \hat{w}(X_i^r)\,\mathbf{1}\{A_i^r=a\}\,Y_i^r}
       {\sum_{i=1}^{n^r} \hat{w}(X_i^r)\,\mathbf{1}\{A_i^r=a\}},
\)
then forms $\hat{\tau}_{\text{ipw}}^o := \hat{\mu}_{+1,\text{ipw}}^o-\hat{\mu}_{-1,\text{ipw}}^o$,
with denominators computed separately within each arm.

In practice, outcome transportability~\eqref{eq:transportability} may fail even when
selection into the trial and treatment assignment are well controlled. This occurs
whenever unmeasured effect modifiers have different distributions across populations.
Our goal is to relax~\eqref{eq:transportability} and develop a sensitivity model that
quantifies how violations of outcome transportability affect conclusions about $\tau^o$.

\subsection{Unmeasured effect modification}
\label{sec:outcome-shift-structural}

To understand when and why outcome transportability~\eqref{eq:transportability} fails,
it helps to consider a structural model for potential outcomes. Suppose
$Y(a) = m_a(X, U, \varepsilon), a \in \{-1, +1\},$
where $U$ is a (possibly multidimensional) unobserved effect moderator and
$\varepsilon$ is idiosyncratic noise. We impose no restriction on the joint
distribution of $(X, U, \varepsilon)$ beyond measurability and integrability,
and we keep the structural functions $m_a$ fixed across populations. Critically,
we allow the conditional distribution of $U$ given $X$ to differ between trial
and target:
$P^r(U \mid X) \neq P^o(U \mid X).$
{This is the key {violation}:} even after conditioning on observed covariates $X$,
the distribution of the unobserved moderator $U$ may differ across populations.

Under Assumption~\ref{assump:internal-validity}, the conditional mean potential
outcome in population $s$ is
$\mu_a^s(x) = \int m_a(x, u, \varepsilon) \, dP^s(u, \varepsilon \mid X = x)$.
Integrating out $\varepsilon$ and defining
$\bar{m}_a(x, u) := \mathbb{E}[m_a(x, u, \varepsilon) \mid X = x, U = u]$, we obtain
$\mu_a^s(x) = \int \bar{m}_a(x, u) \, dP^s(u \mid X = x).$
The \emph{conditional outcome shift} at covariate value $x$ and treatment $a$ is
\begin{equation}
  \Delta_a(x)
  := \mu_a^o(x) - \mu_a^r(x)
  = \int \bar{m}_a(x, u) \, d(P^o - P^r)(u \mid X = x),
  \label{eq:delta-a-def}
\end{equation}
which is nonzero whenever the modifier distribution differs across populations
in directions correlated with $\bar{m}_a(x, \cdot)$.

\paragraph{Confounding versus effect modification.}
It is important to distinguish unmeasured \emph{confounding} (a within-population, treatment-assignment concern) from unmeasured \emph{effect modification} (a cross-population, outcome-distribution concern). Because the trial is randomized, treatment assignment is independent of all covariates---observed and unobserved---within the trial, and we do not assign or observe treatment in the target. Therefore, unmeasured confounding of treatment assignment is irrelevant to our setting.
Unmeasured effect modification, by contrast, can shift the conditional distribution of potential outcomes across populations even after adjusting for $X$.\footnote{For example, consider a trial of an antihypertensive drug with outcome $Y$ (change in systolic blood pressure), observed covariates $X$ (age, sex, BMI, baseline BP), and an unmeasured moderator $U$ (APOE genotype affecting drug metabolism). If APOE-$\varepsilon$4 carrier frequency is ${\sim}15\%$ in the trial but ${\sim}25\%$ in the target population, then $U$ is not a confounder (treatment is randomized) but can still modify the treatment response; this leads to different potential outcome distributions in the trial and target populations, violating outcome transportability.}

A simple example clarifies the mechanics. Suppose the conditional mean is linear
in a scalar $U$:
$
  \mathbb{E}[Y(a) \mid X = x, U = u] = \mu_a^r(x) + \beta_a(x) \cdot u.
$
Then~\eqref{eq:delta-a-def} implies
$
  \Delta_a(x) = \beta_a(x) \bigl( \mathbb{E}^o[U \mid X = x] - \mathbb{E}^r[U \mid X = x] \bigr),
$
an omitted-variable-bias decomposition for external validity \cite{nguyen2018sensitivity, huang2024sensitivity}. Here, the outcome shift is
driven by two factors: the strength of effect modification $\beta_a(x)$, and the
difference in the conditional distribution of $U$ between populations. If either
is zero, transportability holds. But in general, neither $\beta_a(x)$ nor the
shift in $P^s(U \mid X)$ is identified from observed data.

\paragraph{Connecting mean shift to the likelihood-ratio parameter.}
To build intuition for how the sensitivity parameter $\Lambda$ relates to the magnitude of effect modification, consider a Gaussian location-shift model: suppose $Y \mid (a, x, S=r) \sim N(\mu, \sigma^2)$ in the trial and $Y \mid (a, x, S=o) \sim N(\mu + \delta, \sigma^2)$ in the target, both restricted to a bounded interval $[L, U]$. The density ratio is $r(y) = \exp\!\bigl(\delta(y - \mu)/\sigma^2 - \delta^2/(2\sigma^2)\bigr)$, which is monotone in $y$. The maximum over $y \in [L, U]$ gives the required sensitivity parameter:
$\Lambda = \max_{y \in [L,U]} r(y) = \exp\!\bigl(\delta \max(|L-\mu|, |U-\mu|)/\sigma^2 - \delta^2/(2\sigma^2)\bigr).$
Thus, for fixed $(L,U)$ and $\sigma^2$, the required $\Lambda$ increases with the mean shift magnitude $|\delta|$; for fixed $\delta$, wider effective outcome support relative to $\sigma$ yields larger $\Lambda$.

\section{Outcome-Shift Sensitivity Model}
\label{sec:outcome-msm}

Rather than using a sensitivity model that aims to characterize the behavior of the unobserved moderator $U$, we bound the discrepancy between outcome distributions in the trial
and target populations using a likelihood ratio constraint. This parallels
marginal sensitivity models for unmeasured confounding \citep{tan2006marginal,zhao2019sensitivity},
but applies the bound to the outcome rather than treatment assignment.

\subsection{The sensitivity model}
\label{sec:outcome-msm-def}

Let $f^s(y \mid a, x)$ denote the conditional density 
of $Y$ given $(A = a, X = x)$ in population $s \in \{r, o\}$. Under randomization
in the trial, $f^r(y \mid a, x)$ is identified from the observed data $(Y, A, X, S = r)$.

\begin{definition}[Outcome-shift sensitivity model]
\label{def:outcome-msm}
Fix a sensitivity parameter $\Lambda \geq 1$. We say that the pair $(P^r, P^o)$
satisfies the \emph{outcome-shift sensitivity model} with parameter $\Lambda$ if
for each $a \in \{-1, +1\}$ and almost all $x$, the likelihood ratio
\[
  r_a(x, y) := \frac{f^o(y \mid a, x)}{f^r(y \mid a, x)}
\]
satisfies:
\begin{enumerate}[(i)]
  \item (Bounded likelihood ratio)
  $\Lambda^{-1} \leq r_a(x, y) \leq \Lambda \quad \text{for all } y \in \mathcal{Y}$;
  \item (Normalization) For $P_X^r$-almost every $x$,
  $\int_{\mathcal{Y}} r_a(x, y) \, f^r(y \mid a, x) \, dy = 1.$
\end{enumerate}
\end{definition}

The normalization condition holds separately for each covariate value $x$, ensuring that $f^o(\cdot \mid a, x) = r_a(x, \cdot)\, f^r(\cdot \mid a, x)$ is a valid density for each $(a, x)$. Together, the two conditions imply
$f^o(y \mid a, x) = r_a(x, y) \, f^r(y \mid a, x)$ with
$\Lambda^{-1} \leq r_a(x, y) \leq \Lambda$. 
%
The sensitivity parameter $\Lambda$ controls the degree of departure from
transportability. When $\Lambda = 1$, the model enforces
$f^o(\cdot \mid a, x) = f^r(\cdot \mid a, x)$, recovering the standard
transportability assumption~\eqref{eq:transportability}. As $\Lambda$ increases,
progressively larger outcome shifts are permitted. For interpretation, $\Lambda = 2$
means the target population can have at most twice (or at least half) the density
of any outcome value compared to the trial, conditional on $(A, X)$.

\paragraph{Mean-shift implication.}
The LR restriction implies a corresponding bound on conditional mean shifts. For any $(a,x)$,
\[
\begin{aligned}
  |\mu_a^o(x) - \mu_a^r(x)|
  &= \left|\int y \bigl(r_a(x,y)-1\bigr) f^r(y\mid a,x)\,dy\right|
  \le \int |y|\,|r_a(x,y)-1|\,f^r(y\mid a,x)\,dy \\
  &\le (\Lambda-1)\,\mathbb{E}^r\!\left[|Y|\mid A=a,X=x\right],
\end{aligned}
\]
since $r_a(x,y)\in[\Lambda^{-1},\Lambda]$ implies $|r_a(x,y)-1|\le \Lambda-1$. The converse is false: a mean-shift bound alone does not constrain the density ratio, so it permits arbitrary distributional distortions that can yield vacuous worst-case bounds.

\paragraph{Relationship to pointwise likelihood-ratio and f-divergence constraints.}
Our pointwise LR constraint $r_a(x,y) \in [\Lambda^{-1}, \Lambda]$ is an $L^\infty$ constraint on the density ratio, equivalent to bounding the R\'enyi $\infty$-divergence: $D_\infty(P^o \| P^r) \le \log \Lambda$. It is strictly stronger than any single f-divergence constraint: if $r(y) \in [1/\Lambda, \Lambda]$ for all $y$, then for any convex $f$ with $f(1)=0$,
\[
  D_f(P^o \| P^r) \le \max_{t \in [1/\Lambda, \Lambda]} f(t).
\]
The converse is false: an f-divergence bound controls the \emph{average} distortion and can permit large local density-ratio spikes on small-probability sets, whereas our pointwise bound rules these out, yielding tighter sensitivity intervals.
We also note an important distinction from \citet{frauen2023sharp}, whose generalized marginal sensitivity model (GMSM) uses pointwise LR bounds to model unmeasured confounding within a single observational population (internal validity). Our constraint is on outcome distributions across two populations (trial vs.\ target), addressing external validity.

Given $r_a(x, y)$, the conditional mean potential outcome in the target population is
\begin{equation}
  \mu_a^o(x)
  = \int y \, f^o(y \mid a, x) \, dy
  = \int y \, r_a(x, y) \, f^r(y \mid a, x) \, dy
  = \mathbb{E}^r\bigl[Y \cdot r_a(X, Y) \mid A = a, X = x\bigr],
  \label{eq:mu-ao-ratio}
\end{equation}
and the marginal mean is $\mu_a^o = \mathbb{E}^o[\mu_a^o(X)]$. Since $r_a(x, y)$ is
not identified from data, neither is $\mu_a^o$. Our goal is to characterize the
set of values $\mu_a^o$ can take as $r_a$ ranges over all functions satisfying
Definition~\ref{def:outcome-msm}.

To connect this to estimation from trial data, let $w(x) := dP^o / dP^r(x)$ denote
the density ratio between target and trial covariate distributions, and let
$\pi^r(a \mid x)$ denote the (known) trial randomization probability. The target
mean can be written as an expectation over the trial distribution:
\begin{equation}
  \mu_a^o
  = \mathbb{E}^o[\mu_a^o(X)]
  = \mathbb{E}^r\bigl[w(X) \cdot \mu_a^o(X)\bigr]
  = \mathbb{E}^r\biggl[w(X) \cdot r_a(X, Y) \cdot Y \cdot \frac{\mathbf{1}\{A = a\}}{\pi^r(a \mid X)}\biggr].
  \label{eq:mu-ao-weighted}
\end{equation}
This identity is key for estimation: it expresses the target quantity as a weighted
average of trial outcomes, where the weights combine the covariate shift correction
$w(X)$, the outcome shift $r_a(X, Y)$, and the inverse propensity weight
$\mathbf{1}\{A = a\}/\pi^r(a \mid X)$. Since $r_a$ is unknown, we optimize over
admissible likelihood ratios to obtain bounds.

\subsection{Sharp identification bounds}
\label{sec:sharp-bounds}

For a fixed $\Lambda \geq 1$, define the class of admissible likelihood ratio functions
\[
  \mathcal{R}_a(\Lambda)
  := \Bigl\{ r_a : \mathcal{X} \times \mathcal{Y} \to \mathbb{R}_+ \;\Big|\;
     \Lambda^{-1} \leq r_a(x, y) \leq \Lambda, \;
     \mathbb{E}^r[r_a(X, Y) \mid X, A = a] = 1 \text{ a.s.} \Bigr\}.
\]
The identified set for $\mu_a^o$ is
$\mathcal{M}_a(\Lambda) := \{\mu_a^o(r_a) : r_a \in \mathcal{R}_a(\Lambda)\}$,
and the identified set for the target ATE is
$\mathcal{T}(\Lambda) := \{\mu_{+1}^o(r_{+1}) - \mu_{-1}^o(r_{-1}) : r_a \in \mathcal{R}_a(\Lambda)\}$.

To characterize these sets, we first derive bounds on the conditional mean
$\mu_a^o(x)$ for fixed $(x, a)$, then aggregate over the covariate distribution.
Define the optimization problems
\begin{align}
  \mu_a^{o,+}(x; \Lambda)
  &:= \sup_{r_a \in \mathcal{R}_a(\Lambda)}
      \int y \, r_a(x, y) \, f^r(y \mid a, x) \, dy \; \text{ and}
      \label{eq:mu-plus-def} \\
  \mu_a^{o,-}(x; \Lambda)
  &:= \inf_{r_a \in \mathcal{R}_a(\Lambda)}
      \int y \, r_a(x, y) \, f^r(y \mid a, x) \, dy.
      \label{eq:mu-minus-def}
\end{align}

\paragraph{Structure of optimal solutions.}
The optimization problems~\eqref{eq:mu-plus-def}--\eqref{eq:mu-minus-def} have a
special structure: the objective is linear in the likelihood ratio $r_a(x, y)$,
and the constraint set is defined by box constraints plus a single linear equality
(normalization). A standard result in linear programming implies that optimal
solutions to such problems take extreme values---the likelihood ratio equals either
$\Lambda^{-1}$ or $\Lambda$ almost everywhere, rather than intermediate values.

To build intuition, consider the maximization problem for $\mu_a^o(x)$. We seek a
reweighting of the trial outcome distribution that places as much probability mass
as possible on large outcome values. The likelihood ratio $r_a(x, y)$ acts as a
redistributor of probability: setting $r_a(x, y) = \Lambda$ inflates the probability
of outcome $y$ by a factor of $\Lambda$, while $r_a(x, y) = \Lambda^{-1}$ deflates it.
The normalization constraint enforces that total probability remains one, so
inflation at some values must be offset by deflation elsewhere.

Given this trade-off, the optimal strategy is clear: inflate the largest outcomes
as much as possible ($r_a = \Lambda$) and deflate the smallest outcomes as much as
possible ($r_a = \Lambda^{-1}$). Any intermediate value of $r_a$ wastes capacity
that could be better allocated to the extremes. The following lemma formalizes this.

\begin{lemma}[Threshold structure]
\label{lem:threshold}
Fix $x$ and $a$, and suppose $Y \mid (A = a, X = x, S = r)$ has support contained
in a bounded interval $[L, U]$. Any maximizer of~\eqref{eq:mu-plus-def} takes the
threshold form $r_a^\star(x, y) = \Lambda$ for $y > t$, $r_a^\star(x, y) = \Lambda^{-1}$
for $y < t$, and $r_a^\star(x, t) = r_0 \in [\Lambda^{-1}, \Lambda]$, for some
threshold $t \in [L, U]$ determined by the normalization constraint. For the
minimization problem~\eqref{eq:mu-minus-def}, the optimal solution has the reversed
form. The proof is given in Appendix~\ref{app:proof-threshold}.
\end{lemma}

Lemma~\ref{lem:threshold} implies that the conditional identified set for $\mu_a^o(x)$
is the interval $[\mu_a^{o,-}(x; \Lambda), \mu_a^{o,+}(x; \Lambda)]$, with endpoints
attained by threshold likelihood ratios. Aggregating over the covariate distribution
yields sharp bounds on the marginal quantities.

\begin{theorem}[Sharp bounds for the target ATE]
\label{thm:sharp-bounds-pop}
Suppose Assumption~\ref{assump:internal-validity} holds, $|Y| \leq C$ almost surely
in both populations, and $(P^r, P^o)$ satisfies the outcome-shift sensitivity model
with parameter $\Lambda \geq 1$. Then the identified set for the target mean
potential outcome is
\[
  \mathcal{M}_a(\Lambda)
  = \bigl[\mu_a^{o,-}(\Lambda), \, \mu_a^{o,+}(\Lambda)\bigr],
  \quad \text{where} \quad
  \mu_a^{o,\pm}(\Lambda) := \mathbb{E}^o\bigl[\mu_a^{o,\pm}(X; \Lambda)\bigr].
\]
The identified set for the target ATE is
$
  \mathcal{T}(\Lambda)
  = \bigl[\tau^{o,-}(\Lambda), \, \tau^{o,+}(\Lambda)\bigr],
$
with
$
  \tau^{o,-}(\Lambda) := \mu_{+1}^{o,-}(\Lambda) - \mu_{-1}^{o,+}(\Lambda),
  \,
  \tau^{o,+}(\Lambda) := \mu_{+1}^{o,+}(\Lambda) - \mu_{-1}^{o,-}(\Lambda).
$
These bounds are \emph{sharp}: for every value in the interval, there exists a
pair $(P^r, P^o)$ satisfying the sensitivity model and consistent with the observed
trial data for which the target ATE equals that value.
The proof is provided in Appendix \ref{app:proof-theorem}.
\end{theorem}

Theorem~\ref{thm:sharp-bounds-pop} establishes that the outcome-shift sensitivity
model yields a one-dimensional sensitivity analysis indexed by $\Lambda$. When
$\Lambda = 1$, we recover point identification under standard transportability:
$\tau^{o,-}(1) = \tau^{o,+}(1) = \tau^o$. As $\Lambda \to \infty$, the interval
$\mathcal{T}(\Lambda)$ expands toward the range of values consistent with the
trial data and covariate distribution alone, with no restriction on outcome
transportability.

\section{Estimation}
\label{sec:estimation}

We now derive sample analogues of the sharp bounds in Theorem~\ref{thm:sharp-bounds-pop}
and provide an efficient algorithm for computing them.

\subsection{Discrete optimization formulation}
\label{sec:discrete-opt}

Fix a treatment level $a \in \{-1, +1\}$ and let
$\mathcal{I}_a := \{i \in \{1, \ldots, n^r\} : A_i^r = a\}$ denote the indices of
units in the trial receiving treatment $a$, with $n_a := |\mathcal{I}_a|$.
Let $\hat{w}_i := \hat{w}(X_i^r)$ be estimated generalization weights (e.g., from
logistic regression of the study indicator $S$ on covariates $X$ using pooled data).
Define normalized baseline weights
\begin{equation}
  p_i^{(a)} := \frac{\hat{w}_i}{\sum_{j \in \mathcal{I}_a} \hat{w}_j},
  \qquad i \in \mathcal{I}_a.
  \label{eq:baseline-weights}
\end{equation}
These weights sum to one and represent the trial outcome distribution reweighted
to match the target covariate distribution.

In the finite-sample setting, the outcome-shift sensitivity model reduces to
finding multipliers $\lambda_i^{(a)} \in [\Lambda^{-1}, \Lambda]$ such that the
reweighted probabilities $q_i^{(a)} := p_i^{(a)} \lambda_i^{(a)}$ sum to one.
The sample bounds on the target mean solve the linear programs
\begin{equation}
  \hat{\mu}_a^{o,+}(\Lambda) := \max_{\lambda} \sum_{i \in \mathcal{I}_a}
    p_i^{(a)} \lambda_i^{(a)} Y_i^r,
  \qquad
  \hat{\mu}_a^{o,-}(\Lambda) := \min_{\lambda} \sum_{i \in \mathcal{I}_a}
    p_i^{(a)} \lambda_i^{(a)} Y_i^r,
  \label{eq:sample-lp}
\end{equation}
subject to $\Lambda^{-1} \leq \lambda_i^{(a)} \leq \Lambda$ for all $i \in \mathcal{I}_a$
and $\sum_{i \in \mathcal{I}_a} p_i^{(a)} \lambda_i^{(a)} = 1$.
The sample ATE bounds are 
\[
  \hat{\tau}^{o,-}(\Lambda) := \hat{\mu}_{+1}^{o,-}(\Lambda) - \hat{\mu}_{-1}^{o,+}(\Lambda),
  \qquad
  \hat{\tau}^{o,+}(\Lambda) := \hat{\mu}_{+1}^{o,+}(\Lambda) - \hat{\mu}_{-1}^{o,-}(\Lambda).
\]

\subsection{A greedy algorithm}
\label{sec:greedy-alg}

Although~\eqref{eq:sample-lp} is a linear program, its special structure admits a
closed-form solution. The feasible set is a polytope defined by box constraints
plus a single equality constraint (normalization). A standard result in linear
programming implies that extreme points of this polytope have at most one coordinate
strictly between its bounds---all other coordinates are either $\Lambda^{-1}$ or $\Lambda$.
Combined with the threshold structure from Section~\ref{sec:sharp-bounds}, this
means the optimal solution assigns maximum weight $\Lambda$ to the largest outcomes
and minimum weight $\Lambda^{-1}$ to the smallest, with at most one outcome receiving
an intermediate weight to satisfy normalization.

This leads to a simple greedy algorithm. Sort the outcomes in treatment arm $a$ in
ascending order: $Y_{(1)}^r \leq Y_{(2)}^r \leq \cdots \leq Y_{(n_a)}^r$, with
corresponding baseline weights $p_{(1)}^{(a)}, \ldots, p_{(n_a)}^{(a)}$. Initialize
all reweighted probabilities at their lower bound:
$q_{(j)}^{\mathrm{low}} := p_{(j)}^{(a)} \Lambda^{-1}$. The total mass at the lower
bound is $\Lambda^{-1}$, leaving a ``budget'' of $B := 1 - \Lambda^{-1}$ to distribute.
Each outcome $j$ can absorb additional mass up to its capacity
$C_{(j)} := p_{(j)}^{(a)}(\Lambda - \Lambda^{-1})$.

\begin{wrapfigure}[24]{r}{0.52\textwidth}
\vspace{-6pt}
\begin{algorithm2e}[H]
\caption{Outcome-shift sensitivity bounds}
\label{alg:bounds}
\footnotesize
\KwIn{Trial data $\{(X_i^r, A_i^r, Y_i^r)\}_{i=1}^{n^r}$, weights $\hat{w}_i$, $\Lambda \geq 1$}
\KwOut{Bounds $[\hat{\tau}^{o,-}, \hat{\tau}^{o,+}]$ on target ATE}
\For{$a \in \{-1, +1\}$}{
  $\mathcal{I}_a \gets \{i : A_i^r = a\}$, $n_a \gets |\mathcal{I}_a|$\;
  $p_i^{(a)} \gets \hat{w}_i / {\textstyle\sum_{j \in \mathcal{I}_a} \hat{w}_j}$ for $i \in \mathcal{I}_a$\;
  Sort: $Y_{(1)}^r \leq \cdots \leq Y_{(n_a)}^r$ with weights $p_{(j)}^{(a)}$\;
  \For{$j = 1, \ldots, n_a$}{
    $q_{(j)}^{\mathrm{low}} \gets p_{(j)}^{(a)} \Lambda^{-1}$, \enspace
    $C_{(j)} \gets p_{(j)}^{(a)}(\Lambda - \Lambda^{-1})$\;
  }
  $B \gets 1 - \Lambda^{-1}$, \enspace $q^{\max} \gets q^{\mathrm{low}}$\;
  \For{$j = n_a$ \KwTo $1$}{
    $\delta \gets \min\{C_{(j)}, B\}$\;
    $q_{(j)}^{\max} \mathrel{+}= \delta$, \enspace $B \mathrel{-}= \delta$\;
    \lIf{$B = 0$}{\textbf{break}}
  }
  $\hat{\mu}_a^{o,+} \gets \sum_{j=1}^{n_a} q_{(j)}^{\max} Y_{(j)}^r$\;
  $B \gets 1 - \Lambda^{-1}$, \enspace $q^{\min} \gets q^{\mathrm{low}}$\;
  \For{$j = 1$ \KwTo $n_a$}{
    $\delta \gets \min\{C_{(j)}, B\}$\;
    $q_{(j)}^{\min} \mathrel{+}= \delta$, \enspace $B \mathrel{-}= \delta$\;
    \lIf{$B = 0$}{\textbf{break}}
  }
  $\hat{\mu}_a^{o,-} \gets \sum_{j=1}^{n_a} q_{(j)}^{\min} Y_{(j)}^r$\;
}
\Return{$\hat{\tau}^{o,-} = \hat{\mu}_{+1}^{o,-} - \hat{\mu}_{-1}^{o,+}$,\\
\hspace*{2.5em}$\hat{\tau}^{o,+} = \hat{\mu}_{+1}^{o,+} - \hat{\mu}_{-1}^{o,-}$}
\end{algorithm2e}
\vspace{-10pt}
\end{wrapfigure}

To maximize the mean, allocate the budget starting from the largest outcome: fill
$q_{(n_a)}$ to capacity, then $q_{(n_a-1)}$, and so on until the budget is exhausted.
To minimize the mean, allocate starting from the smallest outcome. The following
theorem confirms this procedure is optimal.

\begin{theorem}[Greedy algorithm]
\label{thm:greedy}
Let $Y_{(1)}^r \leq \cdots \leq Y_{(n_a)}^r$ be the sorted outcomes in arm $a$ with
baseline weights $p_{(j)}^{(a)}$. Define $q_{(j)}^{\mathrm{low}} := p_{(j)}^{(a)} \Lambda^{-1}$,
$C_{(j)} := p_{(j)}^{(a)}(\Lambda - \Lambda^{-1})$, and $B := 1 - \Lambda^{-1}$.

\emph{(\textbf{Upper bound})} Set $q_{(j)}^{\max} := q_{(j)}^{\mathrm{low}}$ for all $j$.
For $j = n_a, n_a - 1, \ldots, 1$: set $\delta_{(j)} := \min\{C_{(j)}, B\}$,
update $q_{(j)}^{\max} \leftarrow q_{(j)}^{\max} + \delta_{(j)}$ and
$B \leftarrow B - \delta_{(j)}$; stop when $B = 0$. Then
\(
  \hat{\mu}_a^{o,+}(\Lambda) = \sum_{j=1}^{n_a} q_{(j)}^{\max} Y_{(j)}^r.
\)

\emph{(\textbf{Lower bound})} Reset $B := 1 - \Lambda^{-1}$ and
$q_{(j)}^{\min} := q_{(j)}^{\mathrm{low}}$. For $j = 1, 2, \ldots, n_a$: set
$\delta_{(j)} := \min\{C_{(j)}, B\}$, update
$q_{(j)}^{\min} \leftarrow q_{(j)}^{\min} + \delta_{(j)}$ and
$B \leftarrow B - \delta_{(j)}$; stop when $B = 0$. Then
\(
  \hat{\mu}_a^{o,-}(\Lambda) = \sum_{j=1}^{n_a} q_{(j)}^{\min} Y_{(j)}^r.
\)
The proof is presented in Appendix \ref{app:proof-greedy}.
\end{theorem}

The algorithm requires $O(n_a \log n_a)$ time for sorting plus $O(n_a)$ for the
greedy allocation, giving overall complexity $O(n^r \log n^r)$.
Algorithm~\ref{alg:bounds} summarizes the complete procedure.
We note that ties in outcomes do not affect correctness: when $Y_{(j)} = Y_{(j+1)} = \cdots = Y_{(j+k)}$, any allocation of budget among tied outcomes yields the same objective value, since all share the same outcome value.

\subsection{Consistency}
\label{sec:consistency}

The sample bounds are consistent for their population counterparts under standard
regularity conditions.

\begin{theorem}[Consistency]
\label{thm:consistency}
Suppose Assumption~\ref{assump:internal-validity} holds, $|Y| \leq C$ almost surely,
and the generalization weights satisfy $\hat{w}_i = w(X_i^r) + o_p(1)$ uniformly,
where $w(x) = dP^o / dP^r(x)$ is bounded and strictly positive on the support of
$P^{r,X}$. Then as $n^r, n^o \to \infty$,
\(
  \hat{\mu}_a^{o,\pm}(\Lambda) \xrightarrow{p} \mu_a^{o,\pm}(\Lambda),
  \,
  \hat{\tau}^{o,\pm}(\Lambda) \xrightarrow{p} \tau^{o,\pm}(\Lambda),
\)
for each $a \in \{-1, +1\}$. The sample identified set
$[\hat{\tau}^{o,-}(\Lambda), \hat{\tau}^{o,+}(\Lambda)]$ converges in Hausdorff
distance to the population identified set $[\tau^{o,-}(\Lambda), \tau^{o,+}(\Lambda)]$. The proof is provided in Appendix \ref{app:proof-consistency}.
\end{theorem}


\section{Experiments}
\label{sec:experiments}

We evaluate the proposed bounds via simulation, examining coverage, sharpness, and
comparison to alternatives. We use four data-generating processes (DGPs) with
unmeasured effect modification: DGP~1 (linear Gaussian), DGP~2 (nonlinear),
DGP~3 (binary outcomes), and DGP~4 (heavy-tailed errors). Full DGP specifications,
additional experiments, and extended results appear in Appendix~\ref{app:extended-sim}.

\subsection{Sensitivity Envelopes}
\label{sec:exp-envelopes}

\begin{figure}[t]
\centering
\includegraphics[width=.7\linewidth]{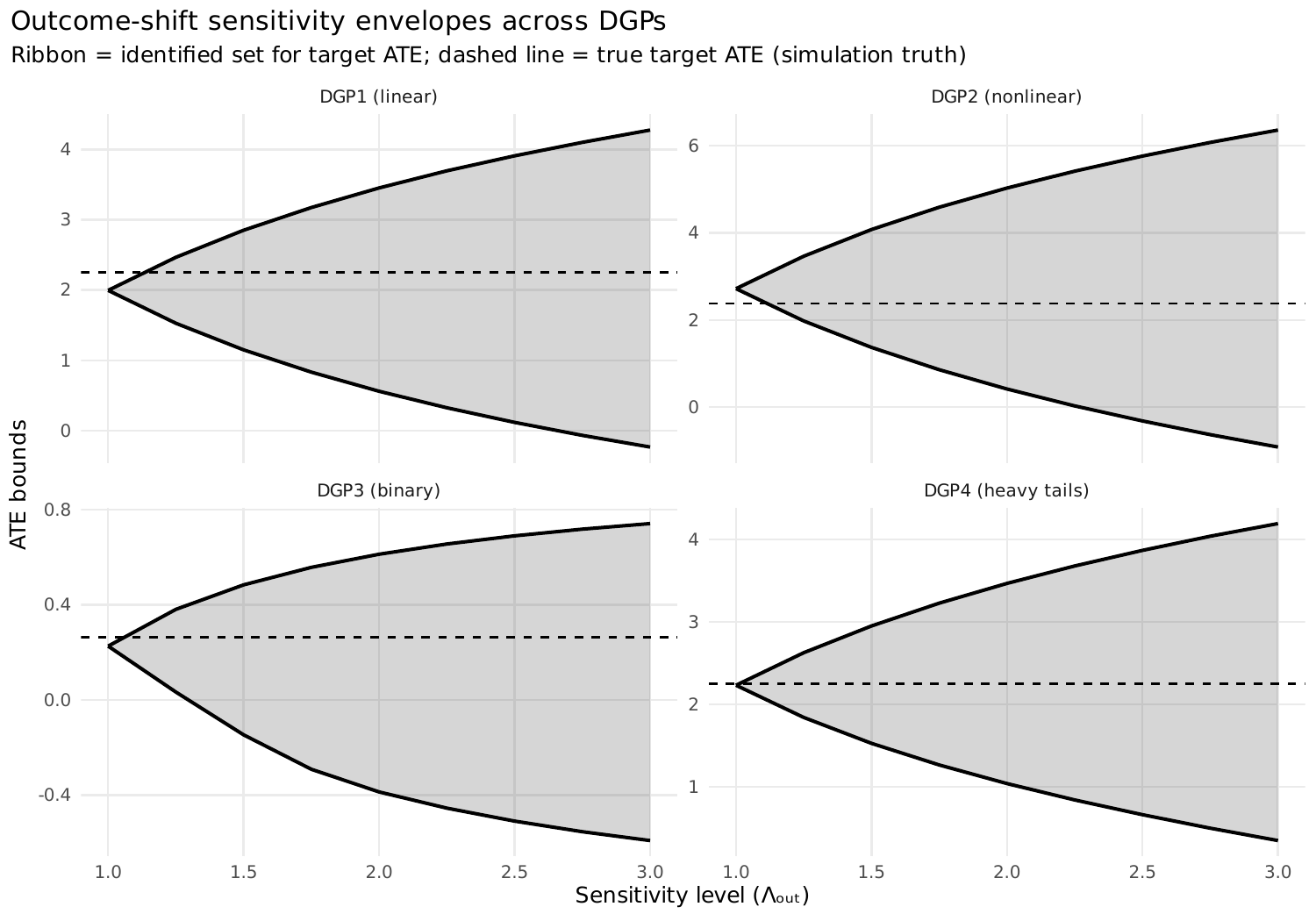}
\caption{Sensitivity envelopes for DGPs 1--4. Each panel shows the sharp bound
interval $[\hat\tau^{o,-}(\Lambda), \hat\tau^{o,+}(\Lambda)]$
as $\Lambda$ varies, with
the true target ATE marked. Binary outcomes (DGP~3) yield tighter bounds.}
\label{fig:envelopes-main}
\end{figure}

Figure~\ref{fig:envelopes-main} displays sensitivity envelopes---the sharp bounds as a
function of $\Lambda$---for a single large dataset ($n^r = 2000$, $n^o = 5000$)
from each DGP. As $\Lambda$ increases from 1 (standard transportability), the
identified set widens monotonically. The true target ATE falls within the bounds 
once $\Lambda$ is sufficiently large. Binary outcomes (DGP~3) yield notably tighter 
intervals due to bounded support, while heavy-tailed errors (DGP~4) produce wider 
intervals for the same $\Lambda$.

\paragraph{Practical interpretation.}
We recommend interpreting the sensitivity envelopes in Figure~\ref{fig:envelopes-main} via three complementary tools: (i)~the \emph{sensitivity curve} itself, which shows how conclusions change as $\Lambda$ increases; (ii)~the \emph{tipping-point} $\Lambda^\star$, defined by
\[
  \Lambda^\star := \inf\{\Lambda \ge 1 : 0 \in [\tau^{o,-}(\Lambda), \tau^{o,+}(\Lambda)]\},
\]
the smallest outcome-shift magnitude at which the population identified set includes the null; and (iii)~\emph{benchmarking against observed modifiers}: when an effect modifier is observed in both populations, one can stratify within treatment arms and compare estimated outcome densities (e.g., via kernel density estimates) between modifier strata to obtain a rough reference scale for plausible $\Lambda$ values (e.g., comparing outcomes in younger vs.\ older trial participants within an arm). We note that density-ratio estimation can be noisy in finite samples (especially in the tails), so such benchmarking should be treated as a heuristic calibration tool.

\paragraph{Two reasons the interval may fail to cover the true ATE.}
Before examining repeated-sampling properties, we clarify two conceptually distinct failure modes. First, if $\Lambda$ is set below the true minimum $\Lambda_{\min}$ (the essential supremum of the actual density ratio between target and trial), the true target distribution lies \emph{outside} the sensitivity model and coverage failure reflects an inherent identification limitation, not a statistical one. Second, even when $\Lambda \ge \Lambda_{\min}$, finite-sample estimation error in outcomes and generalization weights can cause the \emph{estimated} bounds to miss the truth; our consistency result (Theorem~\ref{thm:consistency}) guarantees this error vanishes as $n \to \infty$. Our sensitivity model does not count individual unmeasured modifiers: $\Lambda$ bounds the \emph{aggregate} density ratio shift from all unmeasured sources combined, regardless of whether it arises from one strong modifier or many weak ones.

\subsection{Coverage and Sharpness}
\label{sec:exp-coverage}

\begin{figure}[t]
\centering
\includegraphics[width=0.48\textwidth]{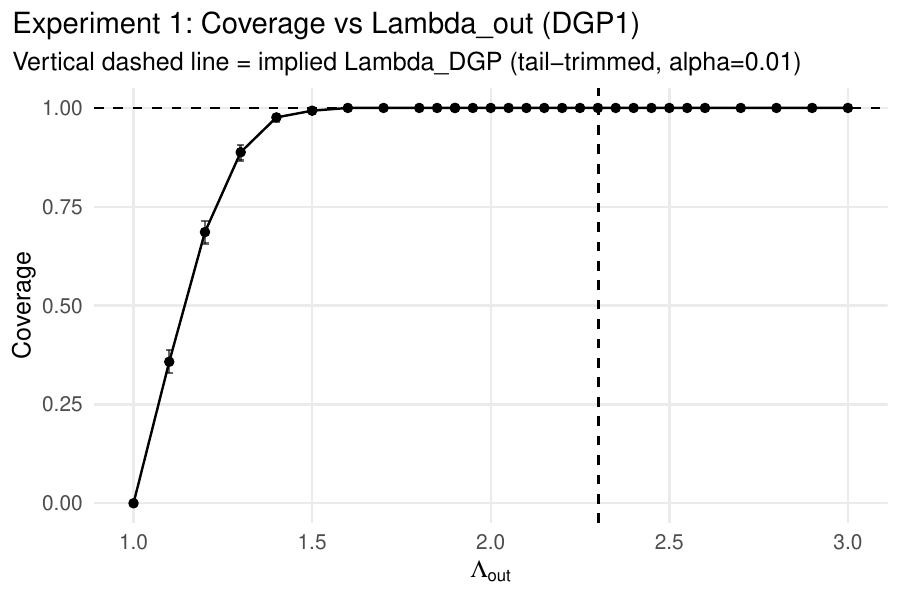}\hfill
\includegraphics[width=0.48\textwidth]{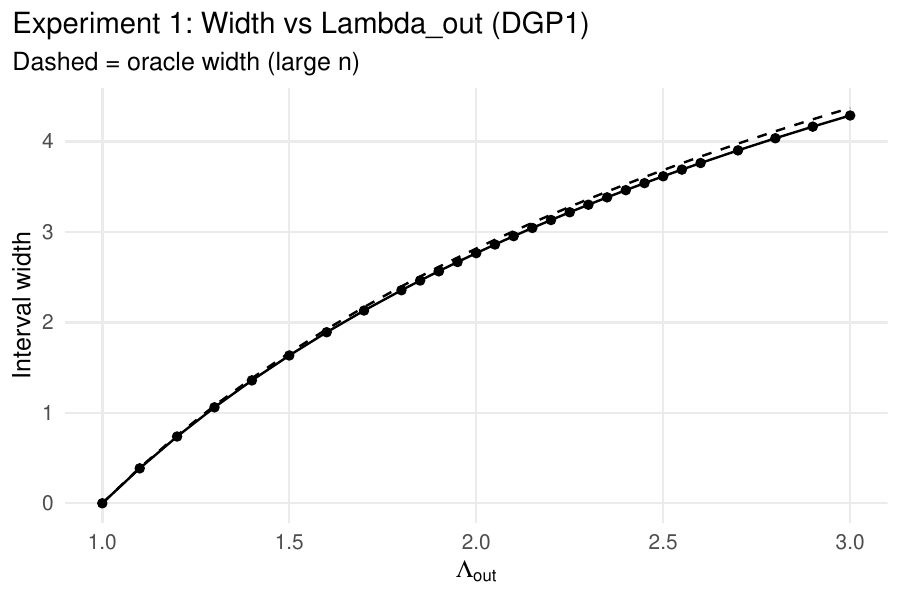}
\caption{Coverage (left) and mean width (right) vs.\ $\Lambda$ for DGP~1 
($R = 1000$). Coverage transitions from undercoverage at $\Lambda = 1$ to nominal 
levels by $\Lambda \approx 1.5$. Oracle width (dashed) confirms sharpness.}
\label{fig:coverage-main}
\end{figure}

We assess repeated-sampling properties using DGP~1 with $(n^r, n^o) = (500, 1000)$
and $R = 1000$ replications. Figure~\ref{fig:coverage-main} shows coverage and mean
interval width as functions of $\Lambda$. At $\Lambda = 1$, coverage is near zero 
because transportability is violated. Coverage increases smoothly, reaching 0.98 at 
$\Lambda = 1.4$ and 1.00 by $\Lambda = 1.6$. The bounds are sharp: the sharpness 
ratio (finite-sample width divided by large-$n$ oracle width) exceeds 0.98 across 
the $\Lambda$ grid (Table~\ref{tab:coverage-main}).

\begin{table}[t]
\centering
\caption{Selected operating points for DGP~1 ($R = 1000$). Sharpness ratio 
confirms finite-sample bounds approximate population bounds.}
\label{tab:coverage-main}
\small
\begin{tabular}{@{}rcccc@{}}
\toprule
$\Lambda$ & Coverage & Mean width & Oracle width & Sharpness \\
\midrule
1.4 & 0.976 & 1.358 & 1.382 & 0.983 \\
1.5 & 0.993 & 1.633 & 1.662 & 0.983 \\
2.0 & 1.000 & 2.764 & 2.815 & 0.982 \\
\bottomrule
\end{tabular}
\end{table}

\subsection{Comparison to Alternatives}
\label{sec:exp-baselines}

We compare our sharp bounds to three alternatives: (i) the naive transported
point estimate ($\Lambda = 1$), (ii) a nonparametric bootstrap CI for the 
transported estimator, and (iii) worst-case bounds assuming only bounded outcomes. 
Figure~\ref{fig:baselines-main} presents results for DGP~1 at $\Lambda = 2.0$ with 
$R = 500$ replications. The naive point estimate achieves 0\% coverage. The bootstrap 
CI achieves approximately 70\% coverage with width 0.80, correctly quantifying sampling 
but not identification uncertainty. Our sharp bounds achieve 100\% coverage with width 
2.76. Worst-case bounds also achieve 100\% coverage but with width 14.6---over five 
times wider---demonstrating the value of the structured sensitivity model.

\subsection{Identification vs.\ Estimation}
\label{sec:exp-identification}

\begin{figure}[t]
\centering
\begin{minipage}[t]{0.32\textwidth}
\centering
\includegraphics[width=\textwidth,trim={0cm 1.5cm 10cm 0cm},clip]{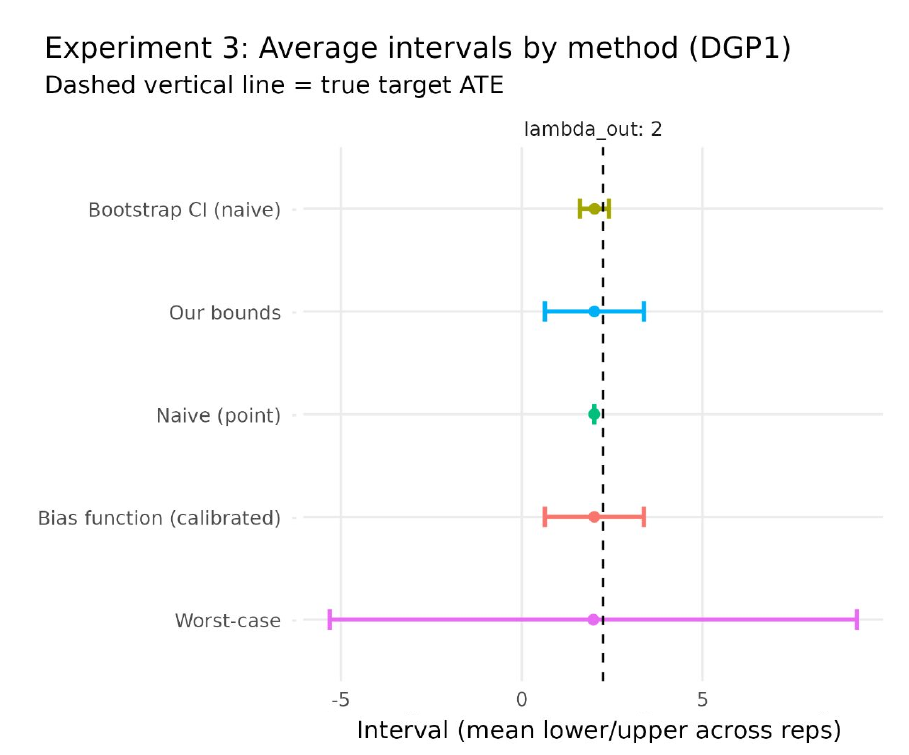}
\captionof{figure}{Comparison at $\Lambda = 2.0$ for DGP~1. Naive estimator and bootstrap 
CI fail to cover; worst-case bounds are uninformatively wide.}
\label{fig:baselines-main}
\end{minipage}\hfill
\begin{minipage}[t]{0.65\textwidth}
\centering
\includegraphics[width=0.48\textwidth]{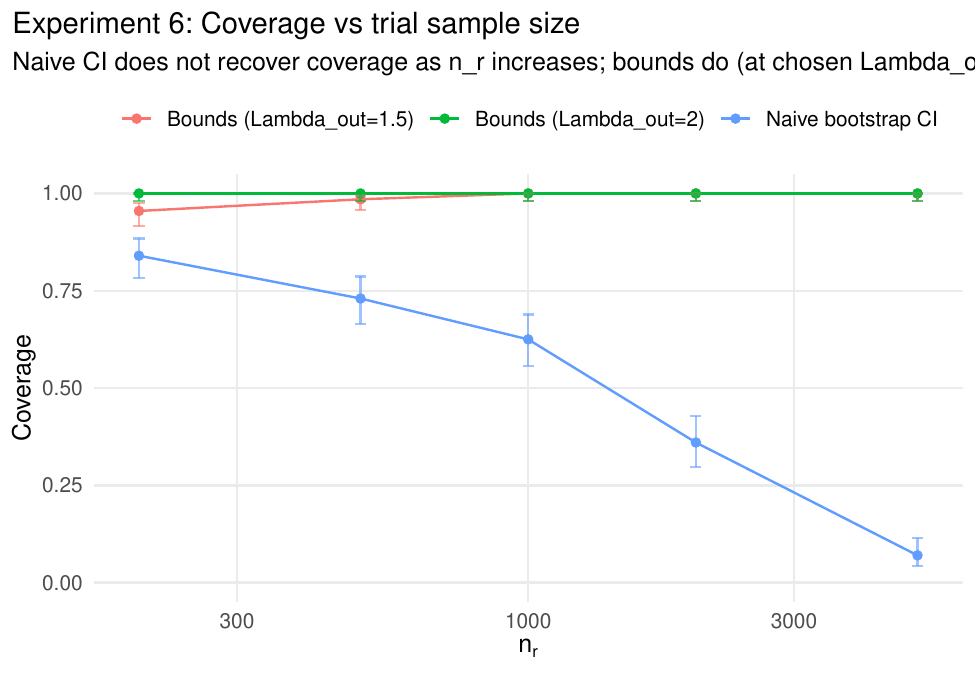}\hfill
\includegraphics[width=0.48\textwidth]{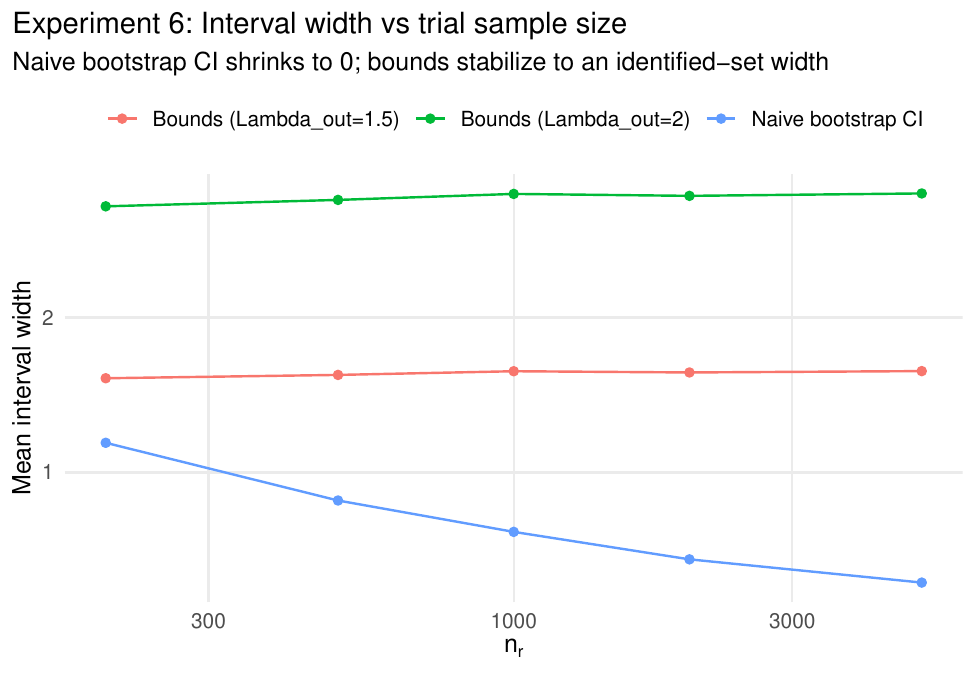}
\captionof{figure}{Identification vs.\ estimation (DGP~1). As $n^r$ grows, naive bootstrap 
CI shrinks but coverage deteriorates (left). Sharp bounds maintain coverage with 
stable widths (right).}
\label{fig:id-vs-est-main}
\end{minipage}
\end{figure}

A key conceptual point is that the limitation under violated transportability is
\emph{identification}, not estimation. We fix DGP~1 with $n^o = 5000$ and vary 
$n^r \in \{200, 500, 1000, 2000, 5000\}$ over $R = 200$ replications.
Figure~\ref{fig:id-vs-est-main} shows a striking pattern: as $n^r$ grows, the naive
bootstrap CI shrinks but coverage \emph{worsens}, dropping from approximately
50\% at $n^r = 200$ to under 10\% at $n^r = 5000$. The estimator becomes
precisely wrong. Our sharp bounds (at $\Lambda = 2.0$) maintain near-perfect 
coverage with widths that stabilize rather than collapse---exactly the behavior 
expected under partial identification.

\subsection{Robustness Across Outcome Types}
\label{sec:exp-robustness}

\begin{wraptable}{r}{0.4\textwidth}
\vspace{-3em}
\centering
\caption{Coverage and width at $\Lambda = 2.0$ across DGPs ($R = 1000$). Complete results are provided in Appendix Table \ref{tab:exp5-robust}.}
\label{tab:robustness-main}
\small
\begin{tabular}{@{}lcc@{}}
\toprule
DGP & Cov. & Width \\
\midrule
1 (Linear) & 1.000 & 2.769 \\
2 (Nonlinear) & 1.000 & 4.260 \\
3 (Binary) & 1.000 & 0.977 \\
4 (Heavy-tailed) & 1.000 & 2.545 \\
\bottomrule
\end{tabular}
\vspace{-1em}
\end{wraptable}

Table~\ref{tab:robustness-main} summarizes coverage and width across all four DGPs
at $\Lambda = 2.0$. The method achieves near-nominal coverage in each case.
Binary outcomes (DGP~3) yield the tightest intervals due to bounded support.
Nonlinear effect modification (DGP~2) and heavy tails (DGP~4) require somewhat
wider intervals, but coverage remains high, suggesting robustness to diverse 
outcome distributions.

\section{Conclusion and Future Work}
\label{sec:conclusion}

We developed a sensitivity analysis framework for generalizing treatment effects when 
transportability may fail due to unmeasured effect modifiers, bounding the likelihood 
ratio between target and trial outcome densities by a parameter $\Lambda$ to obtain 
sharp identified sets. The key insight is that extremal distributions have a threshold 
structure---optimal likelihood ratios take only boundary values $\Lambda^{-1}$ and 
$\Lambda$---enabling a closed-form $O(n \log n)$ greedy algorithm. Simulations confirm 
nominal coverage when $\Lambda$ encompasses the true shift, sharpness in finite samples, 
substantially tighter intervals than worst-case bounds, and correct diagnosis of violated
transportability as an identification (not estimation) problem.

\paragraph{Discussion and future work.}
The present paper establishes identification and consistent estimation of sharp bounds, but does not provide confidence intervals that simultaneously account for sampling variability and partial identification. Because the bound endpoints are functionals of the weighted outcome distribution (resembling weighted tail expectations), they are natural candidates for functional delta-method and bootstrap arguments based on Hadamard differentiability under suitable regularity conditions \citep{shapiro2009lectures}; developing a complete inferential theory that accounts for estimated generalization weights is valuable future work.
Several extensions are also natural. One can allow a \emph{heterogeneous} sensitivity level $\Lambda(a,x)$ to vary by subgroup, retaining convexity because the identification problems are conditional in $(a,x)$; or one can replace the pointwise LR envelope with an \emph{f-divergence} constraint (e.g., a KL ball), which yields a larger ambiguity set (and typically wider bounds) but requires different computation than the greedy algorithm.
Finally, our sharp bounds rely on bounded outcome support; with unbounded outcomes, pointwise LR restrictions can lead to vacuous worst-case means, motivating either tail trimming (as we do in simulations) or additional tail/moment restrictions. Practical performance also depends on weight estimation and overlap: uniform weight convergence is strong under poor overlap, so trimming or normalization is often helpful. A real-data case study applying the sensitivity bounds to a concrete trial generalization problem is an important next step.


\acks{We thank the anonymous reviewers for their constructive feedback, which substantially improved the paper. This work was supported in part by the Patient-Centered Outcomes Research Institute (PCORI) award ME-2023C1-32148. Code is available at \href{https://github.com/AsiaeeLab/marginal-sensitivity-for-treatment-effect-generalization}{github.com/AsiaeeLab/marginal-sensitivity-for-treatment-effect-generalization}.}

\bibliography{ref}

\newpage

\appendix

\section{Proofs}

\subsection{Proof of Lemma~\ref{lem:threshold}}
\label{app:proof-threshold}

Write $P$ for the conditional distributions of $Y \mid (A=a,X=x,S=r)$ on $[L,U]$, so
$dP(y)=f^r(y\mid a,x)\,dy$. Consider the maximization problem; the minimization
case is analogous with the ordering reversed.

Define the feasible set
\[
  \mathcal{C}
  :=
  \Bigl\{
    r:[L,U]\to[\Lambda^{-1},\Lambda]
    \ \Big|\ 
    \int r(y)\,dP(y)=1
  \Bigr\}.
\]
This set is convex and weak-$\star$ compact in $L^\infty(P)$, and the objective
$r\mapsto \int y\,r(y)\,dP(y)$ is linear and continuous, so a maximizer exists.

\paragraph{Step 1 (bang--bang/extreme-point structure).}
We claim that any extreme point of $\mathcal{C}$ takes values in
$\{\Lambda^{-1},\Lambda\}$ $P$-a.s., possibly except on a $P$-null set.
Indeed, if $r\in\mathcal{C}$ satisfies $\Lambda^{-1}<r(y)<\Lambda$ on a set
$E$ with $P(E)>0$, we can choose measurable $E_1,E_2\subseteq E$ with
$P(E_1)=P(E_2)>0$ and define for small $\epsilon>0$,
\[
  r_\pm(y):=r(y)\pm \epsilon(\mathbf 1_{E_1}(y)-\mathbf 1_{E_2}(y)).
\]
For $\epsilon$ small enough, $r_\pm\in\mathcal{C}$ and $r=(r_++r_-)/2$, so $r$
is not extreme.

\paragraph{Step 2 (no inversions $\Rightarrow$ threshold form).}
Let $r$ be a maximizer that is an extreme point, hence $r(y)\in\{\Lambda^{-1},\Lambda\}$
$P$-a.s. Define $H:=\{y:r(y)=\Lambda\}$ and $Lw:=\{y:r(y)=\Lambda^{-1}\}$.
Suppose for contradiction that there exist $y_1<y_2$ with $y_1\in H$ and $y_2\in Lw$,
with both points belonging to sets of positive $P$-mass. Then there exist measurable
sets $E_1\subseteq H$ and $E_2\subseteq Lw$ with $P(E_1)=P(E_2)>0$ and such that
$\sup E_1 < \inf E_2$ (take small neighborhoods and trim).
Define $\tilde r$ by swapping the values on $E_1$ and $E_2$:
\[
  \tilde r(y)
  :=
  \begin{cases}
    \Lambda^{-1}, & y\in E_1,\\
    \Lambda, & y\in E_2,\\
    r(y), & \text{otherwise}.
  \end{cases}
\]
Then $\tilde r\in\mathcal{C}$ because $\int \tilde r\,dP=\int r\,dP$ (the swap
changes the integral by $(\Lambda^{-1}-\Lambda)P(E_1)+(\Lambda-\Lambda^{-1})P(E_2)=0$).
Moreover, the objective strictly increases:
\[
  \int y\,\tilde r(y)\,dP(y)-\int y\,r(y)\,dP(y)
  =
  (\Lambda-\Lambda^{-1})
  \Bigl(\int_{E_2} y\,dP(y)-\int_{E_1} y\,dP(y)\Bigr) \;>\; 0,
\]
since all values in $E_2$ exceed all values in $E_1$. This contradicts optimality.

Therefore, up to $P$-null sets, the set $H=\{r=\Lambda\}$ must be an upper tail:
if $y\in H$ and $y'>y$, then $y'\in H$ $P$-a.s. Hence there exists a threshold
$t\in[L,U]$ such that $r(y)=\Lambda$ for $y>t$ and $r(y)=\Lambda^{-1}$ for $y<t$,
$P$-a.s.

\paragraph{Step 3 (atoms/ties at the threshold).}
If $P(\{t\})=0$, the normalization constraint determines $t$ uniquely (up to null sets).
If $P(\{t\})>0$, the normalization constraint may require assigning an intermediate
value on the atom at $t$: set $r(t)=r_0\in[\Lambda^{-1},\Lambda]$ so that
$\int r\,dP=1$. This yields exactly the stated threshold form. \qed

\subsection{Proof of Theorem~\ref{thm:sharp-bounds-pop}}
\label{app:proof-theorem}

\paragraph{Interval structure.}
Fix $a \in \{-1, +1\}$ and $x \in \mathcal{X}$. By Lemma~\ref{lem:threshold}, the
conditional mean $\mu_a^o(x) = \int y \, r_a(x,y) \, f^r(y \mid a, x) \, dy$ is
maximized and minimized over $\mathcal{R}_a(\Lambda)$ by threshold likelihood ratios,
yielding the interval $[\mu_a^{o,-}(x; \Lambda), \mu_a^{o,+}(x; \Lambda)]$. Since
the map $r_a \mapsto \mu_a^o(x)$ is linear and continuous, and the constraint set
$\mathcal{R}_a(\Lambda)$ is convex and compact (in the weak-$\star$ topology on
$L^\infty$), every value in this interval is attained by some admissible $r_a$.

For the marginal mean, we have
\[
  \mu_a^{o,+}(\Lambda)
  = \mathbb{E}^o[\mu_a^{o,+}(X; \Lambda)]
  = \int \mu_a^{o,+}(x; \Lambda) \, dP^{o,X}(x),
\]
and similarly for $\mu_a^{o,-}(\Lambda)$. Since $|Y| \leq C$ almost surely,
the conditional bounds satisfy $|\mu_a^{o,\pm}(x; \Lambda)| \leq C$ for all $x$,
so the integrals are well-defined. The identified set for $\mu_a^o$ is thus
$\mathcal{M}_a(\Lambda) = [\mu_a^{o,-}(\Lambda), \mu_a^{o,+}(\Lambda)]$.

For the target ATE, note that $\tau^o = \mu_{+1}^o - \mu_{-1}^o$ where $\mu_{+1}^o$
and $\mu_{-1}^o$ can be chosen independently (the likelihood ratios $r_{+1}$ and
$r_{-1}$ are separate). Hence the identified set is
\[
  \mathcal{T}(\Lambda)
  = \bigl\{ \mu_{+1}^o - \mu_{-1}^o : \mu_a^o \in \mathcal{M}_a(\Lambda), \, a \in \{-1, +1\} \bigr\}
  = \bigl[ \mu_{+1}^{o,-} - \mu_{-1}^{o,+}, \, \mu_{+1}^{o,+} - \mu_{-1}^{o,-} \bigr].
\]

\paragraph{Sharpness.}
Fix $a\in\{-1,+1\}$. Let $r_a^+$ and $r_a^-$ be measurable maximizers/minimizers
of the conditional problems (they exist by compactness/continuity as above), so that
$\E^o[\mu_a^o(X;r_a^+)]=\mu_a^{o,+}(\Lambda)$ and
$\E^o[\mu_a^o(X;r_a^-)]=\mu_a^{o,-}(\Lambda)$.

For any $\alpha\in[0,1]$, define the convex combination
\[
  r_{a,\alpha}(x,y) := \alpha r_a^+(x,y) + (1-\alpha) r_a^-(x,y).
\]
Since the constraints defining $\mathcal{R}_a(\Lambda)$ are pointwise bounds and a
linear normalization condition, $r_{a,\alpha}\in\mathcal{R}_a(\Lambda)$ for all
$\alpha\in[0,1]$. Moreover, the induced marginal mean is the corresponding convex
combination:
\[
  \E^o\!\left[\mu_a^o\bigl(X; r_{a,\alpha}\bigr)\right]
  = \alpha\,\mu_a^{o,+}(\Lambda) + (1-\alpha)\,\mu_a^{o,-}(\Lambda).
\]
Hence every $\mu_a^o$ in the interval
$[\mu_a^{o,-}(\Lambda),\mu_a^{o,+}(\Lambda)]$ is attainable by choosing
$\alpha=(\mu_a^o-\mu_a^{o,-})/(\mu_a^{o,+}-\mu_a^{o,-})$ (and any $\alpha$ if the
endpoints coincide).

For the ATE, note $r_{+1}$ and $r_{-1}$ are chosen independently, so any pair
$(\mu_{+1}^o,\mu_{-1}^o)$ in the product of intervals is attainable, and therefore
any difference $\tau=\mu_{+1}^o-\mu_{-1}^o$ in
$[\tau^{o,-}(\Lambda),\tau^{o,+}(\Lambda)]$ is attainable. \qed

\subsection{Proof of Theorem~\ref{thm:greedy}}
\label{app:proof-greedy}

Work in $q$-variables: $q_{(j)} := p_{(j)}^{(a)}\lambda_{(j)}$. The constraints
$\Lambda^{-1}\le \lambda_{(j)}\le \Lambda$ and $\sum_j p_{(j)}^{(a)}\lambda_{(j)}=1$
are equivalent to
\[
  q_{(j)} \in [q_{(j)}^{\mathrm{low}}, q_{(j)}^{\mathrm{high}}],
  \qquad
  \sum_{j=1}^{n_a} q_{(j)} = 1,
\]
where $q_{(j)}^{\mathrm{low}}:=p_{(j)}^{(a)}\Lambda^{-1}$ and
$q_{(j)}^{\mathrm{high}}:=p_{(j)}^{(a)}\Lambda$.
The objective is $\sum_j q_{(j)}Y_{(j)}^r$.

\paragraph{Feasibility of the greedy construction.}
The algorithm starts at $q_{(j)}^{\mathrm{low}}$ (which sums to $\Lambda^{-1}$)
and distributes the remaining mass $\Delta_\Lambda=1-\Lambda^{-1}$ by adding at most
$C_{(j)}=q_{(j)}^{\mathrm{high}}-q_{(j)}^{\mathrm{low}}$ to each coordinate, so the
resulting $q$ stays within the box constraints and sums to one.

\paragraph{Optimality (upper bound).}
Let $q$ be any feasible point. Suppose there exist indices $i<j$ such that
$Y_{(i)}^r < Y_{(j)}^r$, $q_{(i)} > q_{(i)}^{\mathrm{low}}$, and
$q_{(j)} < q_{(j)}^{\mathrm{high}}$. Define
\[
  \varepsilon := \min\{q_{(i)}-q_{(i)}^{\mathrm{low}},\; q_{(j)}^{\mathrm{high}}-q_{(j)}\} > 0,
\]
and set $q'_{(i)}:=q_{(i)}-\varepsilon$, $q'_{(j)}:=q_{(j)}+\varepsilon$, leaving all
other coordinates unchanged. Then $q'$ remains feasible (sum preserved; still within
bounds) and strictly improves the objective:
\[
  \sum_k q'_{(k)}Y_{(k)}^r - \sum_k q_{(k)}Y_{(k)}^r
  = \varepsilon\,(Y_{(j)}^r-Y_{(i)}^r) \;>\; 0.
\]
Therefore, at any maximizer there cannot exist such a pair $(i,j)$: whenever a larger
outcome $j$ has not been filled to its upper bound, all smaller outcomes must sit at
their lower bounds. This forces the ``fill from the top'' structure implemented by the
greedy algorithm, with at most one partially filled coordinate (where the remaining mass
runs out).

\paragraph{Optimality (lower bound).}
The minimization case is identical after reversing the ordering: if a smaller outcome
has not been filled while a larger one has received extra mass above its lower bound,
shifting mass downward decreases the objective. Hence the greedy ``fill from the bottom''
procedure is optimal. \qed

\subsection{Proof of Theorem~\ref{thm:consistency}}
\label{app:proof-consistency}
Fix $a\in\{-1,+1\}$ and $\Lambda\ge 1$. Let $\mathcal I_a=\{i: A_i^r=a\}$.
Define the (arm-specific) weighted empirical measure of outcomes
\[
  \hat P_{n,a} := \sum_{i\in\mathcal I_a} p_i^{(a)}\,\delta_{Y_i^r},
  \qquad
  p_i^{(a)}:=\frac{\hat w_i}{\sum_{j\in\mathcal I_a}\hat w_j}.
\]
Let $P_a^w$ denote the population analogue obtained by reweighting the trial
arm-$a$ distribution by the true density ratio $w(X)=dP^{o,X}/dP^{r,X}(X)$:
for any Borel set $B\subseteq[-B,B]$,
\[
  P_a^w(B)
  :=
  \frac{\E^r\!\big[w(X)\,\mathbf 1\{A=a\}\,\mathbf 1\{Y\in B\}\big]}
       {\E^r\!\big[w(X)\,\mathbf 1\{A=a\}\big]}.
\]
The denominator is strictly positive by positivity of $\P^r(A=a)$ and bounded,
strictly positive $w$ on the support of $P^{r,X}$.

\paragraph{Step 1: $\hat P_{n,a}\Rightarrow P_a^w$ (in a strong one-dimensional sense).}
Let $\tilde p_i^{(a)}:=w(X_i^r)/\sum_{j\in\mathcal I_a}w(X_j^r)$ and
$\tilde P_{n,a}:=\sum_{i\in\mathcal I_a}\tilde p_i^{(a)}\delta_{Y_i^r}$.
Uniform consistency $\sup_i|\hat w_i-w(X_i^r)|=o_p(1)$ and boundedness of $w$
imply
\[
  \sum_{i\in\mathcal I_a}\big|p_i^{(a)}-\tilde p_i^{(a)}\big|=o_p(1),
  \qquad\text{hence}\qquad
  d_{\mathrm{TV}}(\hat P_{n,a},\tilde P_{n,a})=o_p(1).
\]
Moreover, since the weights $w(X)$ are bounded and the class
$\{\mathbf 1\{y\le t\}: t\in\R\}$ is VC, a weighted Glivenko--Cantelli theorem yields
\[
  \sup_{t\in\R}\Big|\tilde P_{n,a}((-\infty,t]) - P_a^w((-\infty,t])\Big|
  \xrightarrow{p} 0.
\]
Because $Y$ is supported on $[-B,B]$, uniform convergence of CDFs implies
convergence in $W_1$ (Wasserstein-1) distance, and therefore
$W_1(\hat P_{n,a},P_a^w)\xrightarrow{p}0$.

\paragraph{Step 2: continuity of the sharp-bound functionals.}
For any probability measure $P$ supported on $[-B,B]$, define
\[
  \phi_\Lambda^{+}(P)
  := \sup_{\lambda:\ \Lambda^{-1}\le \lambda \le \Lambda,\ \int \lambda\,dP=1}
      \int y\,\lambda(y)\,dP(y),
  \qquad
  \phi_\Lambda^{-}(P)
  := \inf_{\lambda:\ \Lambda^{-1}\le \lambda \le \Lambda,\ \int \lambda\,dP=1}
      \int y\,\lambda(y)\,dP(y).
\]
The sample LP in \eqref{eq:sample-lp} is exactly
$\hat\mu_a^{o,\pm}(\Lambda)=\phi_\Lambda^{\pm}(\hat P_{n,a})$, and the population
sharp bounds satisfy $\mu_a^{o,\pm}(\Lambda)=\phi_\Lambda^{\pm}(P_a^w)$.

Let $Q_P(u):=\inf\{y: P((-\infty,y])\ge u\}$ be the generalized quantile function
and set $\rho:=1/(\Lambda+1)$. A standard rearrangement/knapsack argument (the
continuous analogue of Theorem~\ref{thm:greedy}) gives the representation
\[ 
  \phi_\Lambda^{+}(P)
  = \Lambda^{-1}\int y\,dP(y)
    +(\Lambda-\Lambda^{-1})\int_{1-\rho}^{1} Q_P(u)\,du,
  \quad
  \phi_\Lambda^{-}(P)
  = \Lambda^{-1}\int y\,dP(y)
    +(\Lambda-\Lambda^{-1})\int_{0}^{\rho} Q_P(u)\,du,
\]
where the generalized quantile handles atoms and corresponds to allowing one
partial weight at the threshold.

In one dimension,
$W_1(P,Q)=\int_0^1 |Q_P(u)-Q_Q(u)|\,du$, and also
$|\int y\,dP(y)-\int y\,dQ(y)|\le W_1(P,Q)$ for measures on a bounded interval.
Therefore,
\[
  |\phi_\Lambda^{\pm}(P)-\phi_\Lambda^{\pm}(Q)|
  \le \Lambda\,W_1(P,Q),
\]
so $\phi_\Lambda^{\pm}$ is continuous (indeed Lipschitz) in $W_1$.

Combining with Step 1 yields
$\hat\mu_a^{o,\pm}(\Lambda)=\phi_\Lambda^{\pm}(\hat P_{n,a})
\xrightarrow{p}\phi_\Lambda^{\pm}(P_a^w)=\mu_a^{o,\pm}(\Lambda)$.

\paragraph{Step 3: ATE bounds and Hausdorff convergence.}
By definition,
$\hat\tau^{o,-}(\Lambda)=\hat\mu_{+1}^{o,-}(\Lambda)-\hat\mu_{-1}^{o,+}(\Lambda)$
and
$\hat\tau^{o,+}(\Lambda)=\hat\mu_{+1}^{o,+}(\Lambda)-\hat\mu_{-1}^{o,-}(\Lambda)$,
so convergence of the means implies
$\hat\tau^{o,\pm}(\Lambda)\xrightarrow{p}\tau^{o,\pm}(\Lambda)$.
Finally, for intervals on $\R$ the Hausdorff distance equals the maximum endpoint
error, hence
$$d_H([\hat\tau^{o,-}(\Lambda),\hat\tau^{o,+}(\Lambda)],
      [\tau^{o,-}(\Lambda),\tau^{o,+}(\Lambda)])\xrightarrow{p}0$$.
\qed

\section{Extended Simulation Study}
\label{app:extended-sim}

This section empirically validates the outcome-shift sensitivity bounds, illustrates
the coverage--informativeness tradeoff as the sensitivity parameter varies, and
documents robustness to nonlinearity, discrete outcomes, heavy tails, and weight
estimation.

\subsection{Estimand, data structure, and sensitivity parameter}
\label{sec:sim-setup}

We consider a randomized trial conducted in a \emph{trial population} (study indicator
$S=r$) and a \emph{target population} ($S=o$). In the trial we observe i.i.d.\ draws
$\{(X_i, A_i, Y_i)\}_{i=1}^{n^r}$, where $A\in\{-1,+1\}$ is randomized with
$\mathbb{P}(A=+1)=1/2$, and in the target we observe baseline covariates
$\{X_j\}_{j=1}^{n^o}$ (no outcomes).

The target estimand is the target average treatment effect (ATE)
\[
\tau^o \;=\; \mathbb{E}\!\left[\,Y^o(+1)-Y^o(-1)\,\right],
\]
where $Y^o(a)$ denotes the potential outcome under treatment level $a$ in the target population.

\paragraph{Outcome-shift sensitivity model and $\Lambda_{\mathrm{out}}$.}
The outcome-shift model in \cref{sec:outcome-msm,def:outcome-msm} constrains how much
the \emph{conditional outcome distribution} in the target can differ from the trial.
In our simulation code and figures we denote the sensitivity parameter by
$\Lambda_{\mathrm{out}}\ge 1$ (this is the same $\Lambda$ as in the main text, with an
``out'' subscript only to emphasize outcome shift). Concretely, for each treatment arm
$a\in\{-1,+1\}$ and covariate profile $x$, we bound the conditional likelihood ratio by
\[
\frac{1}{\Lambda_{\mathrm{out}}}
\;\le\;
\frac{f^o_{Y\mid A,X}(y\mid a,x)}{f^r_{Y\mid A,X}(y\mid a,x)}
\;\le\;
\Lambda_{\mathrm{out}}
\qquad \text{for all } y,
\]
where $f^s_{Y\mid A,X}$ is the conditional outcome density/mass in population $s\in\{r,o\}$.
When $\Lambda_{\mathrm{out}}=1$, this reduces to standard outcome transportability and
the bounds collapse to the usual transported point estimate.

\paragraph{Generalization weights.}
To transport trial outcomes to the target covariate distribution we use inverse odds
weights (see \cref{eq:baseline-weights}). In most experiments we use \emph{oracle}
weights $w(x)=f_X^o(x)/f_X^r(x)$ (available in simulation) to isolate outcome-shift
effects from weight-estimation error. We separately study estimated weights in
\cref{sec:sim-exp6}.

\paragraph{Metrics.}
Across Monte Carlo replications we report:
(i) \emph{coverage} of the true $\tau^o$ by the bound interval;
(ii) \emph{mean interval width};
(iii) a \emph{sharpness ratio} defined as (sample mean width)/(oracle width),
where ``oracle width'' is the width obtained by running the same procedure on an
extremely large simulated trial to approximate the population sharp interval; and
(iv) runtime where relevant. For coverage curves we include Wilson binomial intervals
(where plotted; see tables for corresponding $[\,\text{lo},\text{hi}\,]$).

\begin{table}[t]
\centering
\caption{Simulation design summary (master table; can be trimmed).}
\label{tab:sim-design-summary}
\begin{tabular}{@{}llrrrrl@{}}
\toprule
Exp. & Goal & DGP(s) & $n^r$ & $n^o$ & Reps & Key outputs \\
\midrule
Main & Envelopes vs.\ $\Lambda_{\mathrm{out}}$ & 1--4 & 2000 & 5000 & 1 dataset & Fig.~\ref{fig:main-envelopes} \\
1 & Validation + sharpness & 1 & 500 & 1000 & 1000 & Fig.~\ref{fig:exp1-val}, Tab.~\ref{tab:exp1-key} \\
2 & Tradeoff vs.\ moderator shift & 1 & 500 & 1000 & 1000 & Fig.~\ref{fig:exp2-tradeoff}, Tab.~\ref{tab:exp2-breakeven} \\
3 & Baseline comparison & 1 & 500 & 1000 & 50 & Fig.~\ref{fig:exp3-baselines}, Tab.~\ref{tab:exp3-baselines} \\
4 & Scaling in $n^r$ (fixed $\Lambda_{\mathrm{out}}$) & 1 & 100--5000 & 1000 & 300 & Fig.~\ref{fig:exp4-scaling} \\
5 & Robustness across outcome types & 1--4 & 500 & 1000 & 1000 & Fig.~\ref{fig:exp5-robust}, Tab.~\ref{tab:exp5-robust} \\
6 & Weight estimation sensitivity & 1 & 500 & 1000 & 150 & Fig.~\ref{fig:exp6-weights} \\
6b & Identification vs estimation & 1 & 200--5000 & 5000 & 200 & Fig.~\ref{fig:exp6-id-vs-est} \\
7 & Bounded-support DGP (LR holds literally) & 7 & 500 & 1000 & 200 & Fig.~\ref{fig:exp7-bounded} \\
8 & ``Bang--bang'' optimizer structure & 1 & 1 dataset & -- & -- & Fig.~\ref{fig:exp8-bangbang} \\
\bottomrule
\end{tabular}
\end{table}

\subsection{Data-generating processes (DGPs 1--4)}
\label{sec:sim-dgps}

All DGPs share a covariate shift between trial and target and an unmeasured moderator
whose distribution differs across populations. Let $p=5$ and write $X=(X_1,\dots,X_p)^\top$.
We generate
\[
X^r \sim \mathcal{N}(0_p, I_p),
\qquad
X^o \sim \mathcal{N}(\mu_{\mathrm{shift}}\cdot \mathbf{1}_p, I_p),
\]
with $\mu_{\mathrm{shift}}=0.5$ in the main experiments. The unmeasured moderator satisfies
\[
U^s \mid X \sim \mathcal{N}(\gamma_s X_1, 1),
\qquad s\in\{r,o\},
\]
and we set $\gamma_r=0$ and (unless otherwise stated) $\gamma_o=0.5$, so that the target
has a different distribution of the effect modifier even after conditioning on $X$.

\paragraph{DGP 1 (linear outcome with Gaussian moderator).}
Potential outcomes are
\[
Y(a) \;=\; \beta_0 + \beta_x^\top X + a\cdot\bigl(\tau_0 + \beta_u U\bigr) + \varepsilon_a,
\qquad \varepsilon_a \sim \mathcal{N}(0,\sigma^2),
\]
with $\beta_0=1$, $\beta_x=(0.3,0.2,0.1,0.1,0.1)^\top$, $\tau_0=1$, $\beta_u=0.5$, and $\sigma=1$.
Because $A\in\{-1,+1\}$, the individual treatment effect is
$Y(+1)-Y(-1)=2(\tau_0+\beta_u U)$, hence the target ATE has a closed form:
\[
\tau^o \;=\; 2\tau_0 + 2\beta_u\,\mathbb{E}^o[U]
\;=\; 2\tau_0 + 2\beta_u\,\gamma_o\,\mathbb{E}^o[X_1]
\;=\; 2\tau_0 + 2\beta_u\,\gamma_o\,\mu_{\mathrm{shift}}
\;=\; 2 + \gamma_o \mu_{\mathrm{shift}}
\;=\; 2.25.
\]

\paragraph{DGP 2 (nonlinear effect modification).}
Covariates and $U$ are generated as in DGP 1, but the outcome model is nonlinear:
\[
Y(a) \;=\; \beta_0 + \sin(\pi X_1/2) + X_2^2
\;+\; a\cdot\Bigl(\tau_0 + \beta_u\,|U|\cdot \mathrm{sign}(X_1)\Bigr) + \varepsilon_a,
\qquad \varepsilon_a \sim \mathcal{N}(0,\sigma^2).
\]
This preserves the same covariate shift and moderator shift but induces nonlinear
treatment effect heterogeneity. The true $\tau^o$ is approximated via Monte Carlo with
$n_{\text{truth}}=10^6$ draws from the target joint distribution of $(X,U)$.

\paragraph{DGP 3 (binary outcomes).}
We generate binary outcomes via
\[
Y(a)\sim \mathrm{Bernoulli}(p_a(X,U)),
\qquad
\mathrm{logit}\bigl(p_a(X,U)\bigr)=\beta_0 + \beta_x^\top X + a\cdot(\tau_0+\beta_u U),
\]
with $\beta_0=-0.5$, $\beta_x=(0.3,0.2,0.1,0.1,0.1)^\top$, $\tau_0=0.5$, and $\beta_u=0.3$.
The true $\tau^o$ is approximated via Monte Carlo with $n_{\text{truth}}=10^6$ draws.

\paragraph{DGP 4 (heavy-tailed outcomes).}
This matches DGP 1 except the error is heavy-tailed:
\[
Y(a)=\beta_0 + \beta_x^\top X + a\cdot(\tau_0+\beta_u U) + \varepsilon_a,
\qquad \varepsilon_a \sim t_3 \text{ scaled to variance } \sigma^2,
\]
so that rare large residuals occur. The closed-form $\tau^o$ remains the same as DGP 1.

\subsection{Main envelope figure across DGPs}
\label{sec:sim-main-envelope}

\begin{figure}[t]
\centering
\includegraphics[width=\linewidth]{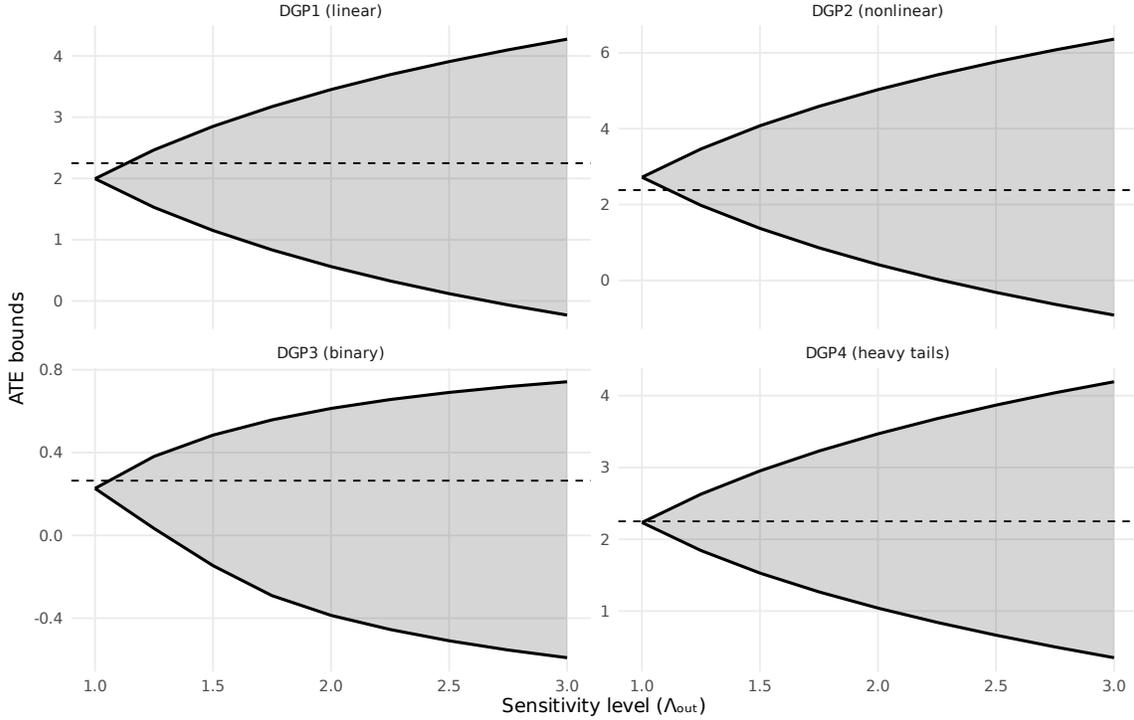}
\caption{Sensitivity envelopes on a single large simulated dataset for DGPs 1--4.
Each panel plots the sharp bound interval $[\hat\tau^-(\Lambda_{\mathrm{out}}),\hat\tau^+(\Lambda_{\mathrm{out}})]$
as a function of $\Lambda_{\mathrm{out}}$, with the true target ATE indicated for reference.
DGP 3 (binary outcomes) yields notably tighter envelopes due to bounded outcomes.}
\label{fig:main-envelopes}
\end{figure}

Before turning to repeated-sampling experiments, we plot a single-dataset ``sensitivity
envelope'' for each DGP: the sharp lower and upper bounds as $\Lambda_{\mathrm{out}}$
varies, using a large trial ($n^r=2000$) and target covariate sample ($n^o=5000$).
\Cref{fig:main-envelopes} shows the expected monotone widening of the identified set
as the analyst allows larger outcome shift. For bounded outcomes (DGP 3), widths are
substantially smaller; for heavy tails (DGP 4), widths grow faster with $\Lambda_{\mathrm{out}}$.

\subsection{Experiment 1: validation and sharpness (DGP 1)}
\label{sec:sim-exp1}

\paragraph{Design.}
We simulate DGP 1 with $(n^r,n^o)=(500,1000)$ for $R=1000$ replications. For each replication
and each value on a grid $\Lambda_{\mathrm{out}}\in[1,3]$, we compute sharp bounds and record:
coverage of $\tau^o=2.25$, mean width, and a sharpness ratio relative to an ``oracle width''
computed on an extremely large simulated dataset ($n^r=100{,}000$) using the same procedure.

Because Gaussian likelihood ratios are technically unbounded, we also compute a
\emph{tail-trimmed} implied $\Lambda_{\mathrm{DGP}}(\alpha)$ by restricting to central
$(1-\alpha)$ outcome mass (here $\alpha=0.01$) and taking the smallest $\Lambda$ that
upper-bounds the conditional likelihood ratio on that truncated region. This provides a
calibration reference (conservative for functionals like the ATE).

\begin{figure}[t]
\centering
\begin{minipage}{0.49\linewidth}
\centering
\includegraphics[width=\linewidth]{img/exp1_dgp1_coverage_vs_lambda.pdf}
\end{minipage}\hfill
\begin{minipage}{0.49\linewidth}
\centering
\includegraphics[width=\linewidth]{img/exp1_dgp1_width_vs_lambda.pdf}
\end{minipage}

\vspace{0.5em}

\begin{minipage}{0.70\linewidth}
\centering
\includegraphics[width=\linewidth]{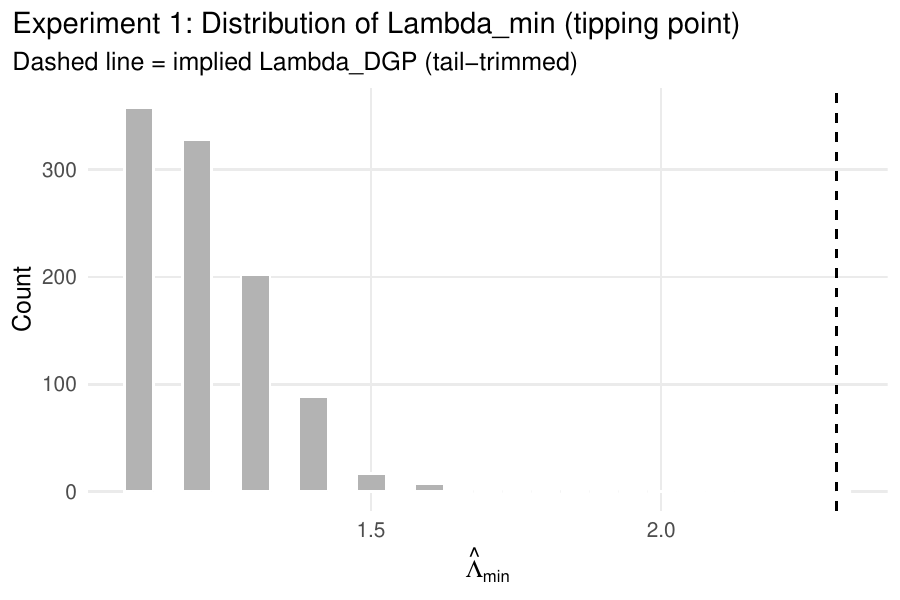}
\end{minipage}

\caption{Experiment 1 (DGP 1): validation and sharpness.
\emph{Top-left:} coverage of the sharp interval for $\tau^o$ vs.\ $\Lambda_{\mathrm{out}}$ (with binomial uncertainty).
\emph{Top-right:} mean interval width vs.\ $\Lambda_{\mathrm{out}}$, overlaid with an oracle large-$n$ width.
\emph{Bottom:} distribution of $\Lambda_{\min}$, the smallest sensitivity level for which a replicate’s interval covers $\tau^o$.}
\label{fig:exp1-val}
\end{figure}

\paragraph{Results.}
\Cref{fig:exp1-val} shows a clean transition from severe undercoverage near
$\Lambda_{\mathrm{out}}=1$ (transportability assumed) to near-nominal coverage once
$\Lambda_{\mathrm{out}}$ is moderately above 1. Coverage is already $\approx 0.98$ at
$\Lambda_{\mathrm{out}}=1.4$ and $\approx 0.99$ at $\Lambda_{\mathrm{out}}=1.5$, reaching
$1.00$ by $\Lambda_{\mathrm{out}}=1.6$. Importantly, widths closely track oracle widths:
the sharpness ratio is about $0.98$ across the grid (Tab.~\ref{tab:exp1-key}),
indicating the finite-sample procedure is nearly as tight as the population sharp interval.
The tipping point distribution in \Cref{fig:exp1-val} (rightmost panel) concentrates around
$\Lambda_{\min}\in[1.1,1.3]$, the smallest sensitivity level needed for a given replicate to cover.

\begin{table}[t]
\centering
\caption{Experiment 1 (DGP 1): selected operating points. Coverage is across $R=1000$
replications; oracle widths use a large-$n$ approximation.}
\label{tab:exp1-key}
\begin{tabular}{@{}rcccc@{}}
\toprule
$\Lambda_{\mathrm{out}}$ & Coverage & Mean width & Oracle width & Sharpness ratio \\
\midrule
1.2 & 0.686 & 0.738 & 0.750 & 0.983 \\
1.4 & 0.976 & 1.358 & 1.382 & 0.983 \\
1.5 & 0.993 & 1.633 & 1.662 & 0.983 \\
1.6 & 1.000 & 1.890 & 1.924 & 0.983 \\
2.0 & 1.000 & 2.764 & 2.815 & 0.982 \\
3.0 & 1.000 & 4.285 & 4.369 & 0.981 \\
\bottomrule
\end{tabular}
\end{table}

\subsection{Experiment 2: coverage--informativeness tradeoff vs.\ moderator shift (DGP 1)}
\label{sec:sim-exp2}

\paragraph{Design.}
We vary the magnitude of population shift in the unmeasured moderator by setting
$\gamma_o\in\{0.25,0.5,0.75,1.0\}$ (with $\gamma_r=0$), while keeping
$(n^r,n^o)=(500,1000)$ and $R=1000$ replications. For each $\gamma_o$ we sweep
$\Lambda_{\mathrm{out}}$ and summarize the induced tradeoff between coverage and width.
We also compute a \emph{breakeven} sensitivity level, defined as the smallest
$\Lambda_{\mathrm{out}}$ achieving at least $95\%$ empirical coverage for that $\gamma_o$.

\begin{figure}[t]
\centering
\begin{minipage}{0.62\linewidth}
\centering
\includegraphics[width=\linewidth]{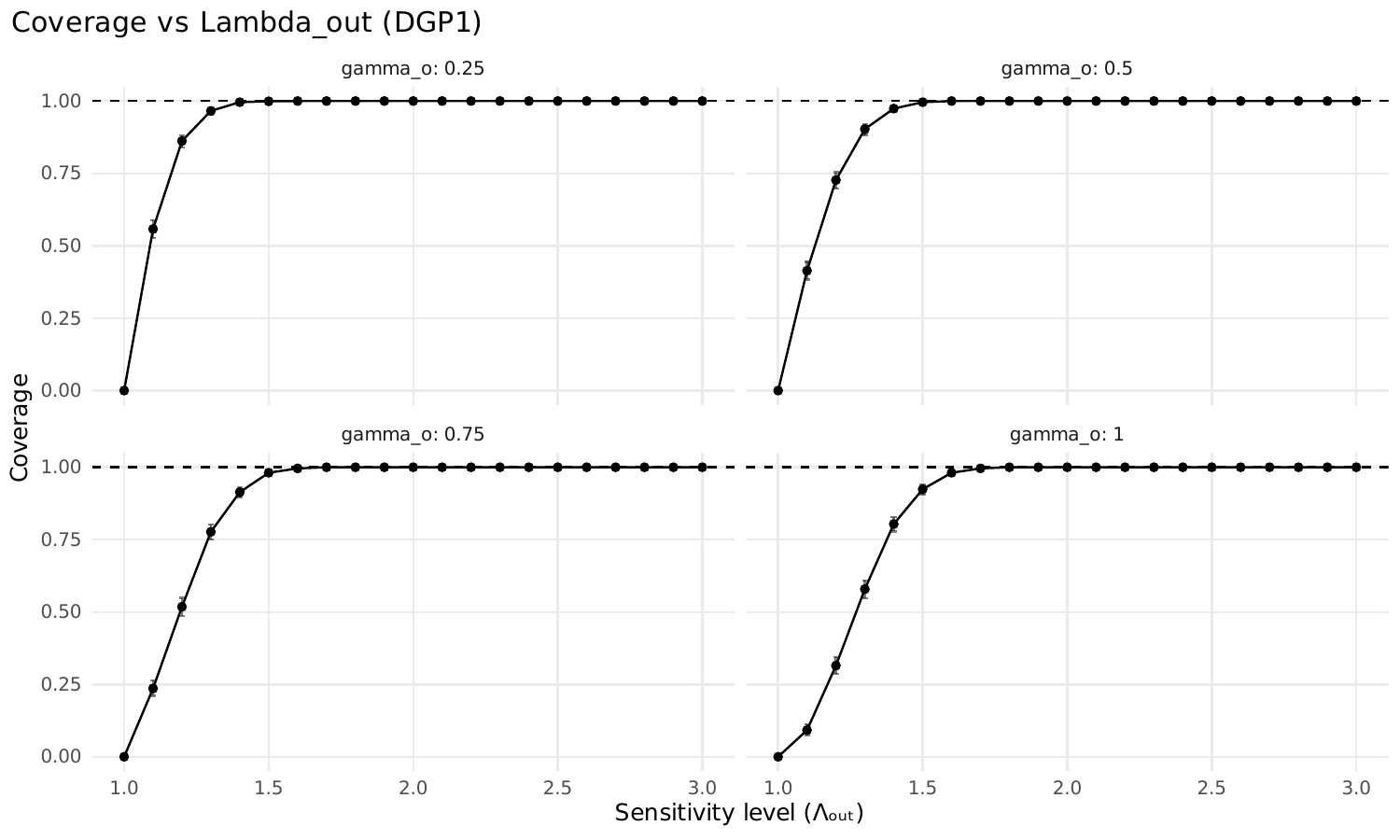}
\end{minipage}\hfill
\begin{minipage}{0.36\linewidth}
\centering
\includegraphics[width=\linewidth]{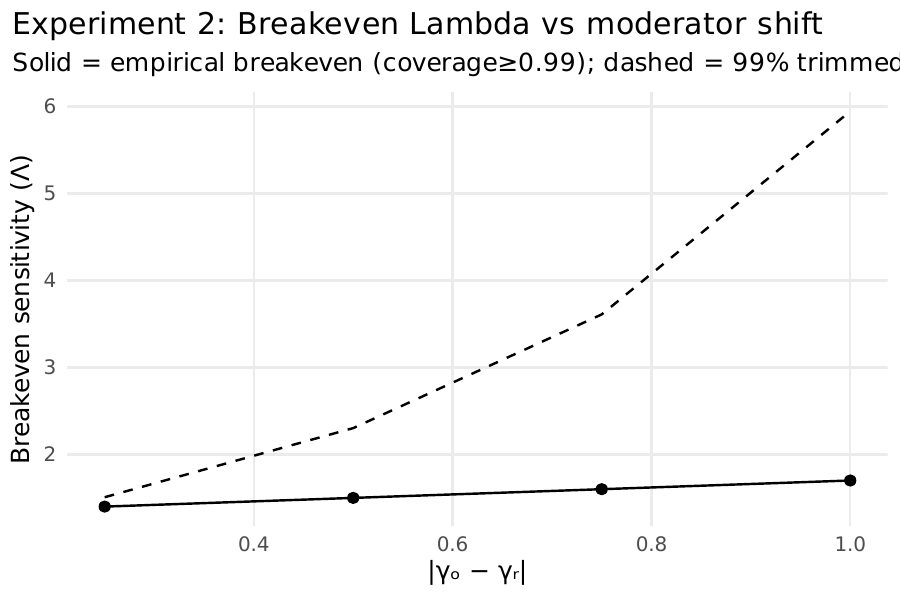}
\end{minipage}

\caption{Experiment 2 (DGP 1): increasing moderator shift $\gamma_o$ pushes the coverage-vs-$\Lambda_{\mathrm{out}}$
curve to the right. \emph{Left:} coverage vs.\ $\Lambda_{\mathrm{out}}$ for $\gamma_o\in\{0.25,0.5,0.75,1.0\}$.
\emph{Right:} breakeven $\Lambda_{\mathrm{out}}$ achieving 95\% coverage.}
\label{fig:exp2-tradeoff}
\end{figure}

\paragraph{Results.}
As expected, increasing moderator shift makes transportability more fragile: the
coverage curve shifts right, and the breakeven $\Lambda_{\mathrm{out}}$ rises nearly
linearly with $\gamma_o$ (Tab.~\ref{tab:exp2-breakeven}, \Cref{fig:exp2-tradeoff}).
Concretely, the breakeven level increases from $1.4$ at $\gamma_o=0.25$ to $1.7$ at
$\gamma_o=1.0$. This experiment provides an interpretable calibration: analysts can
read off the sensitivity level needed to restore nominal coverage under a given
degree of moderator shift.

\begin{table}[t]
\centering
\caption{Experiment 2 (DGP 1): breakeven $\Lambda_{\mathrm{out}}$ (minimum achieving $\ge 0.95$ coverage)
as a function of moderator-shift strength $\gamma_o$. We also report the tail-trimmed
$\Lambda_{\mathrm{DGP}}(\alpha{=}0.01)$ computed from the conditional likelihood ratio, which is
substantially more conservative for the ATE functional.}
\label{tab:exp2-breakeven}
\begin{tabular}{@{}rcc@{}}
\toprule
$\gamma_o$ & Breakeven $\Lambda_{\mathrm{out}}$ (95\% cov.) & Tail-trimmed $\Lambda_{\mathrm{DGP}}(0.01)$ \\
\midrule
0.25 & 1.4 & 1.508 \\
0.50 & 1.5 & 2.302 \\
0.75 & 1.6 & 3.609 \\
1.00 & 1.7 & 5.944 \\
\bottomrule
\end{tabular}
\end{table}

\subsection{Experiment 3: comparison to baseline procedures (DGP 1)}
\label{sec:sim-exp3}

\paragraph{Design.}
We compare our sharp bounds to common alternatives at two sensitivity levels
$\Lambda_{\mathrm{out}}\in\{1.5,2.0\}$ on DGP 1 with $(n^r,n^o)=(500,1000)$. The baselines are:
(i) naive transported point estimate (assumes $\Lambda_{\mathrm{out}}=1$);
(ii) naive nonparametric bootstrap CI for the transported estimator;
(iii) a very conservative worst-case bound; and
(iv) a calibrated bias-function baseline.
(For computational convenience in this experiment, the run uses $R=50$ replications; the
gaps are large enough to be visually stable, and increasing $R$ is straightforward.)

\paragraph{Results.}
The naive transported estimator (and its bootstrap CI) can be too optimistic:
its intervals are narrow but under-cover, while worst-case bounds cover trivially but are
far too wide. Our bounds achieve near-nominal coverage with substantially smaller width
than worst-case, and they coincide with the calibrated bias-function baseline here
(Tab.~\ref{tab:exp3-baselines}).

\begin{table}[t]
\centering
\caption{Experiment 3 (DGP 1): baseline comparison at $\Lambda_{\mathrm{out}}\in\{1.5,2.0\}$.
Coverage is across $R=50$ replications.}
\label{tab:exp3-baselines}
\begin{tabular}{@{}lccc@{}}
\toprule
Method & $\Lambda_{\mathrm{out}}$ & Coverage & Mean width \\
\midrule
Naive (point, $\Lambda_{\mathrm{out}}{=}1$) & 1.5 & 0.00 & 0.00 \\
Bootstrap CI (naive) & 1.5 & 0.70 & 0.804 \\
Our bounds & 1.5 & 0.98 & 1.616 \\
Worst-case & 1.5 & 1.00 & 14.576 \\
\midrule
Naive (point, $\Lambda_{\mathrm{out}}{=}1$) & 2.0 & 0.00 & 0.00 \\
Bootstrap CI (naive) & 2.0 & 0.70 & 0.804 \\
Our bounds & 2.0 & 1.00 & 2.734 \\
Worst-case & 2.0 & 1.00 & 14.576 \\
\bottomrule
\end{tabular}
\end{table}

\begin{figure}[t]
\centering
\includegraphics[width=0.9\linewidth]{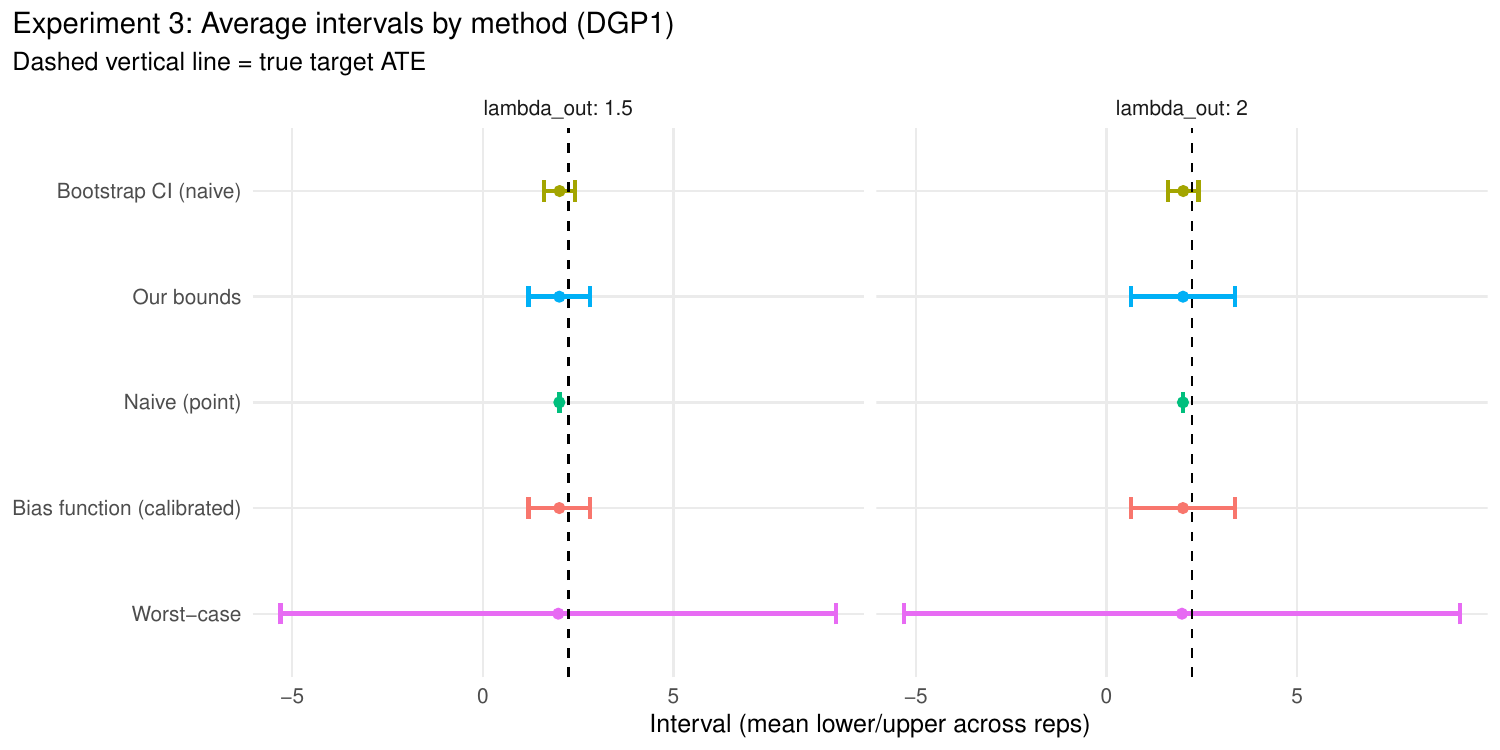}
\caption{Experiment 3 (DGP 1): forest plot comparing interval procedures at
$\Lambda_{\mathrm{out}}\in\{1.5,2.0\}$. Our bounds remain informative while maintaining
high coverage; worst-case bounds are orders of magnitude wider.}
\label{fig:exp3-baselines}
\end{figure}

\subsection{Experiment 4: scaling with trial sample size (DGP 1)}
\label{sec:sim-exp4}

\paragraph{Design.}
Fix $\Lambda_{\mathrm{out}}=2.0$ and $(n^o=1000)$, vary the trial size
$n^r\in\{100,200,500,1000,2000,5000\}$, and run $R=300$ replications under DGP 1.
We record coverage, mean width, sharpness ratio (relative to a large-$n$ oracle width),
and runtime.

\begin{figure}[t]
\centering
\begin{minipage}{0.32\linewidth}
\centering
\includegraphics[width=\linewidth]{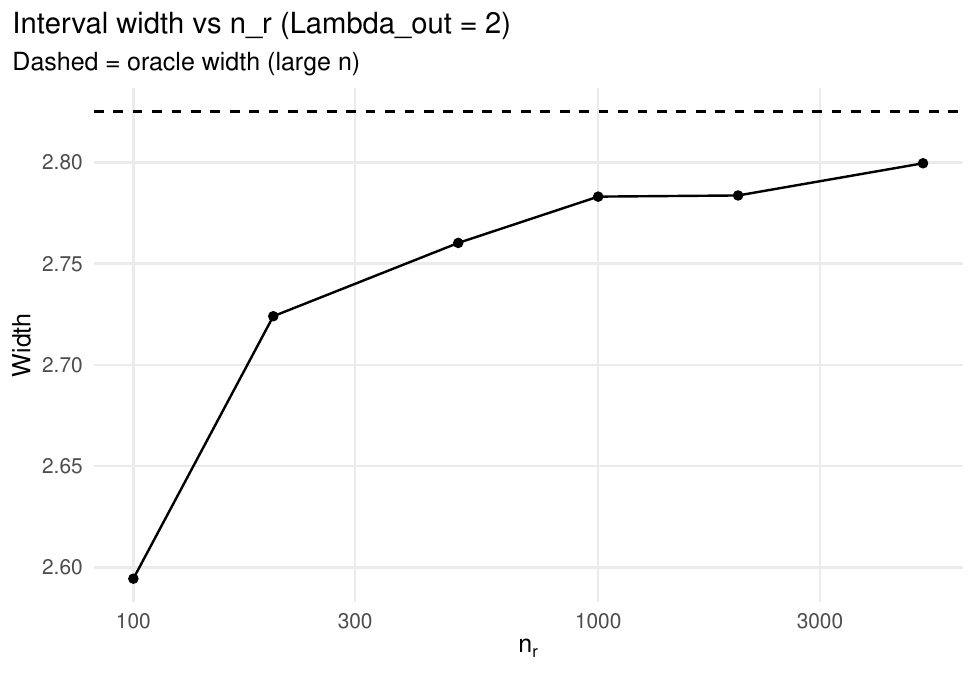}
\end{minipage}\hfill
\begin{minipage}{0.32\linewidth}
\centering
\includegraphics[width=\linewidth]{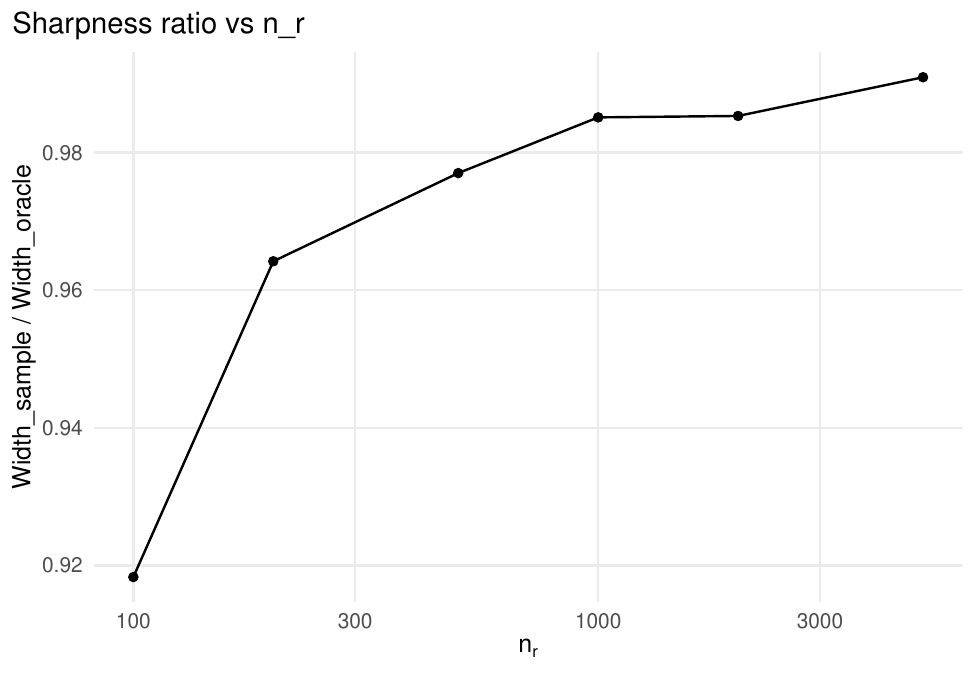}
\end{minipage}\hfill
\begin{minipage}{0.32\linewidth}
\centering
\includegraphics[width=\linewidth]{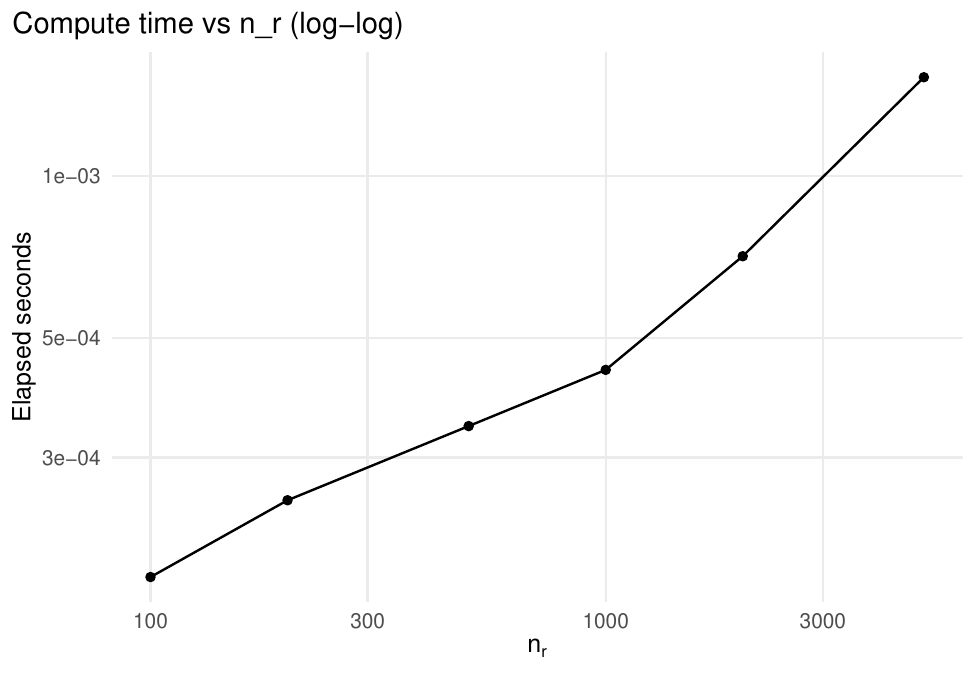}
\end{minipage}
\caption{Experiment 4 (DGP 1): scaling with trial size $n^r$ at fixed $\Lambda_{\mathrm{out}}=2$.
\emph{Left:} mean width vs.\ $n^r$. \emph{Center:} sharpness ratio vs.\ $n^r$ (approaches 1).
\emph{Right:} runtime vs.\ $n^r$ (milliseconds).}
\label{fig:exp4-scaling}
\end{figure}

\paragraph{Results.}
\Cref{fig:exp4-scaling} shows that width and sharpness stabilize quickly as $n^r$ grows,
while runtime remains negligible (milliseconds). Even at $n^r=200$, coverage is already
at 1.00 in this setting, and the sharpness ratio approaches 1 as finite-sample noise
shrinks. This supports the practical usability of the greedy algorithm in
\cref{alg:bounds}.

\subsection{Experiment 5: robustness across outcome types (DGPs 1--4)}
\label{sec:sim-exp5}

\paragraph{Design.}
We run the same Monte Carlo experiment across DGPs 1--4 with $(n^r,n^o)=(500,1000)$,
$R=1000$ replications, and $\Lambda_{\mathrm{out}}\in\{1.5,2.0,3.0\}$.

\begin{figure}[t]
\centering
\includegraphics[width=0.75\linewidth]{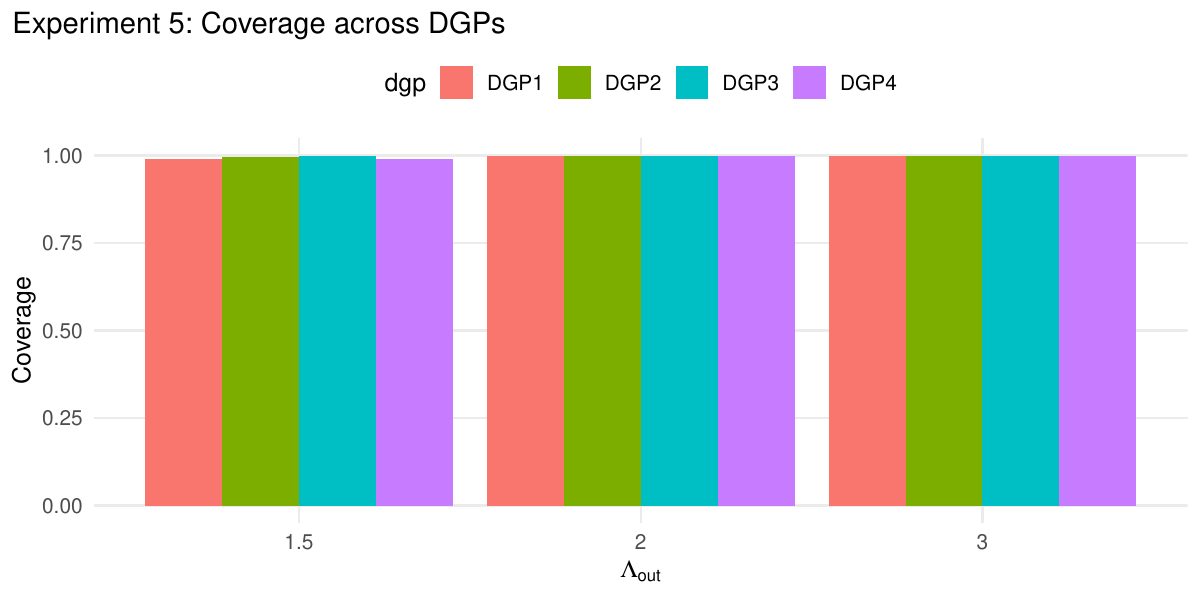}
\caption{Experiment 5: robustness across DGPs 1--4. Coverage at $\Lambda_{\mathrm{out}}=1.5$
is already near nominal in all cases and reaches 1.00 by $\Lambda_{\mathrm{out}}=2.0$ for all DGPs.}
\label{fig:exp5-robust}
\end{figure}

\paragraph{Results.}
\Cref{fig:exp5-robust} and Tab.~\ref{tab:exp5-robust} show that the bounds maintain
high coverage across nonlinear, binary, and heavy-tailed settings. Binary outcomes
(DGP 3) yield the tightest intervals, while nonlinear and heavy-tailed outcomes require
wider intervals for the same $\Lambda_{\mathrm{out}}$, as expected.

\begin{table}[t]
\centering
\caption{Experiment 5: coverage and width across DGPs.}
\label{tab:exp5-robust}
\begin{tabular}{@{}lrrr@{}}
\toprule
DGP & $\Lambda_{\mathrm{out}}$ & Coverage & Mean width \\
\midrule
DGP1 (linear) & 1.5 & 0.991 & 1.635 \\
DGP1 (linear) & 2.0 & 1.000 & 2.769 \\
DGP1 (linear) & 3.0 & 1.000 & 4.296 \\
\midrule
DGP2 (nonlinear) & 1.5 & 0.995 & 2.498 \\
DGP2 (nonlinear) & 2.0 & 1.000 & 4.260 \\
DGP2 (nonlinear) & 3.0 & 1.000 & 6.717 \\
\midrule
DGP3 (binary) & 1.5 & 0.999 & 0.609 \\
DGP3 (binary) & 2.0 & 1.000 & 0.977 \\
DGP3 (binary) & 3.0 & 1.000 & 1.332 \\
\midrule
DGP4 (heavy tails) & 1.5 & 0.989 & 1.497 \\
DGP4 (heavy tails) & 2.0 & 1.000 & 2.545 \\
DGP4 (heavy tails) & 3.0 & 1.000 & 3.991 \\
\bottomrule
\end{tabular}
\end{table}

\subsection{Experiment 6: sensitivity to weight estimation (DGP 1)}
\label{sec:sim-exp6}

\paragraph{Design.}
We compare three weighting strategies in DGP 1 with $(n^r,n^o)=(500,1000)$ and $R=150$ replications:
(i) oracle weights using the known Gaussian density ratio; (ii) estimated logistic
membership weights using the correct covariates; and (iii) a misspecified logistic
membership model using only $X_1$. We report coverage and width at
$\Lambda_{\mathrm{out}}\in\{1.5,2.0\}$ and visualize the induced weight distributions.

\begin{figure}[t]
\centering
\begin{minipage}{0.49\linewidth}
\centering
\includegraphics[width=\linewidth]{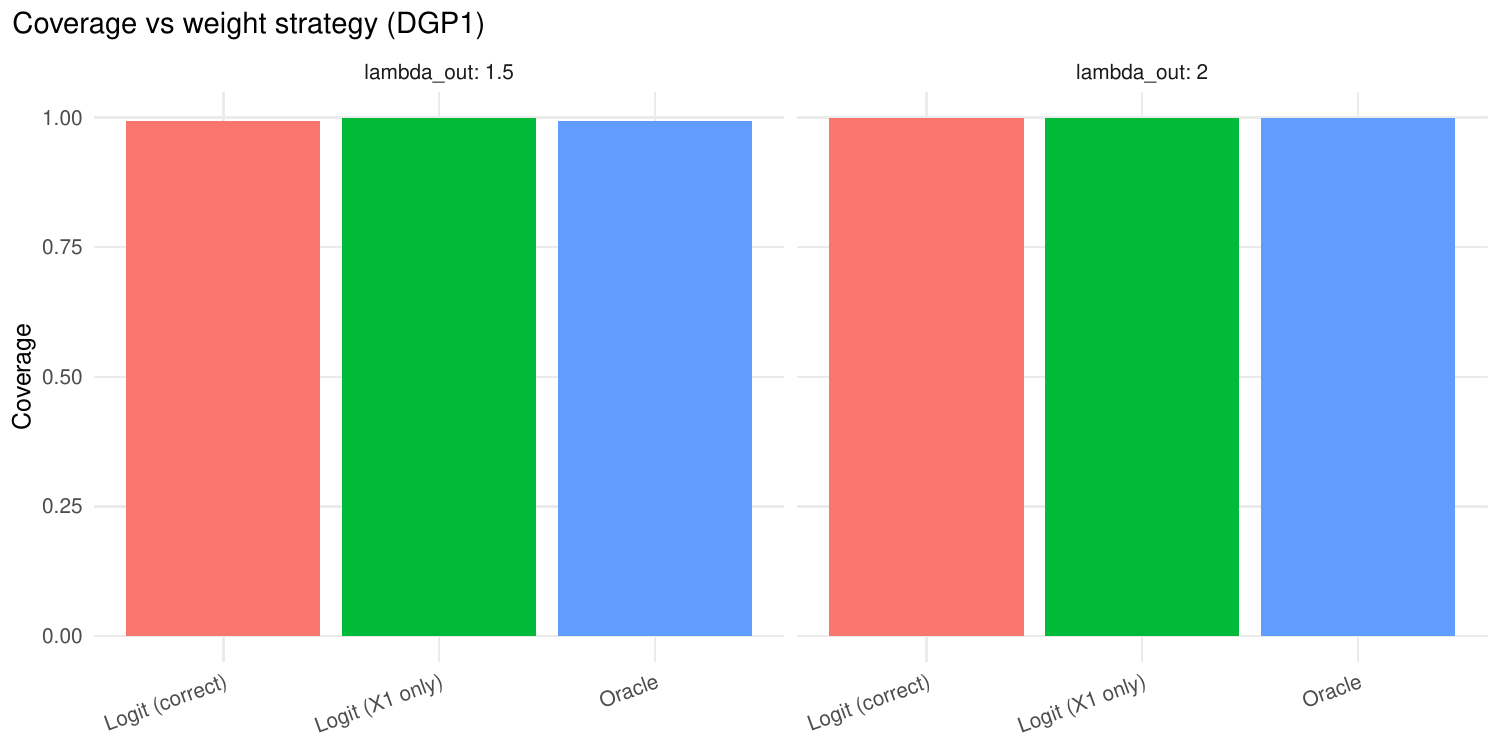}
\end{minipage}\hfill
\begin{minipage}{0.49\linewidth}
\centering
\includegraphics[width=\linewidth]{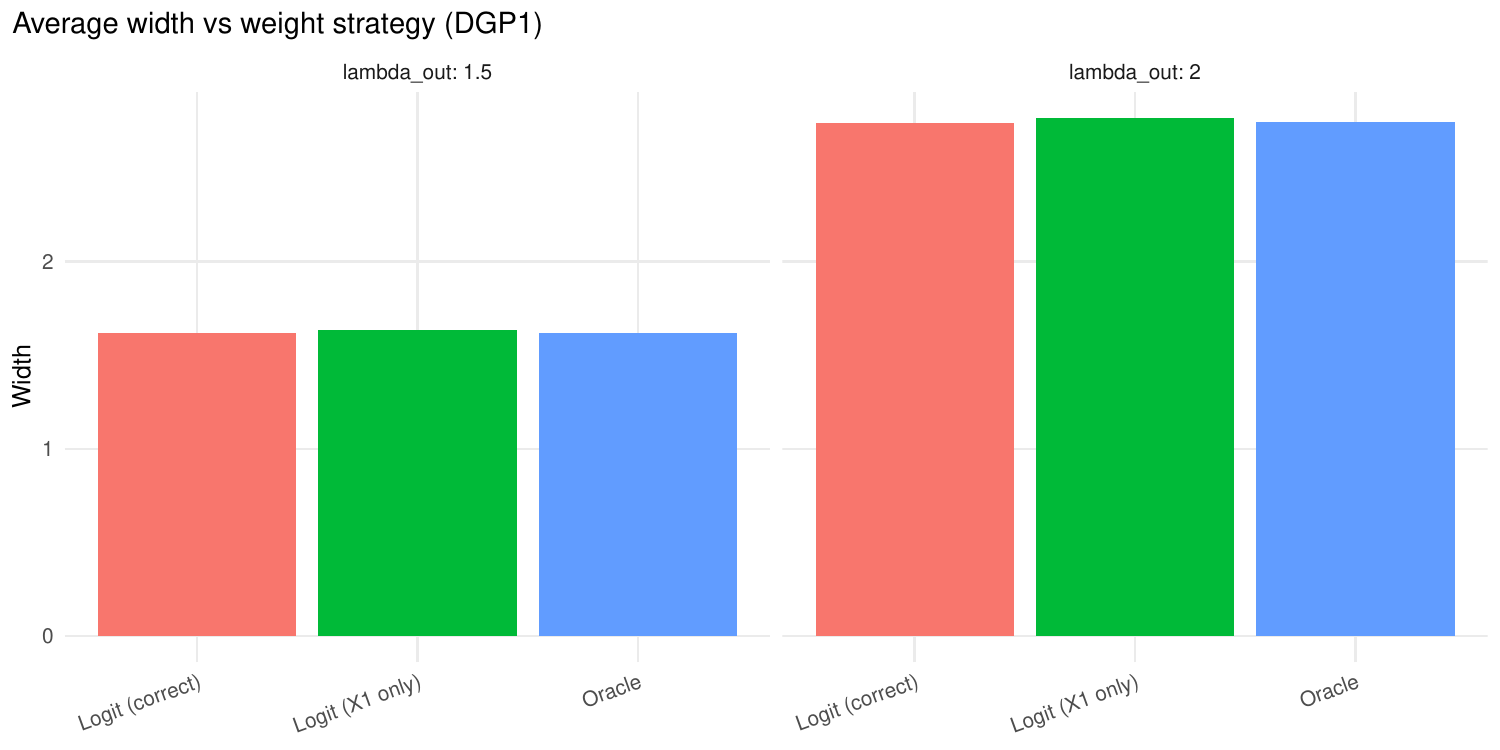}
\end{minipage}

\vspace{0.5em}

\begin{minipage}{0.70\linewidth}
\centering
\includegraphics[width=\linewidth]{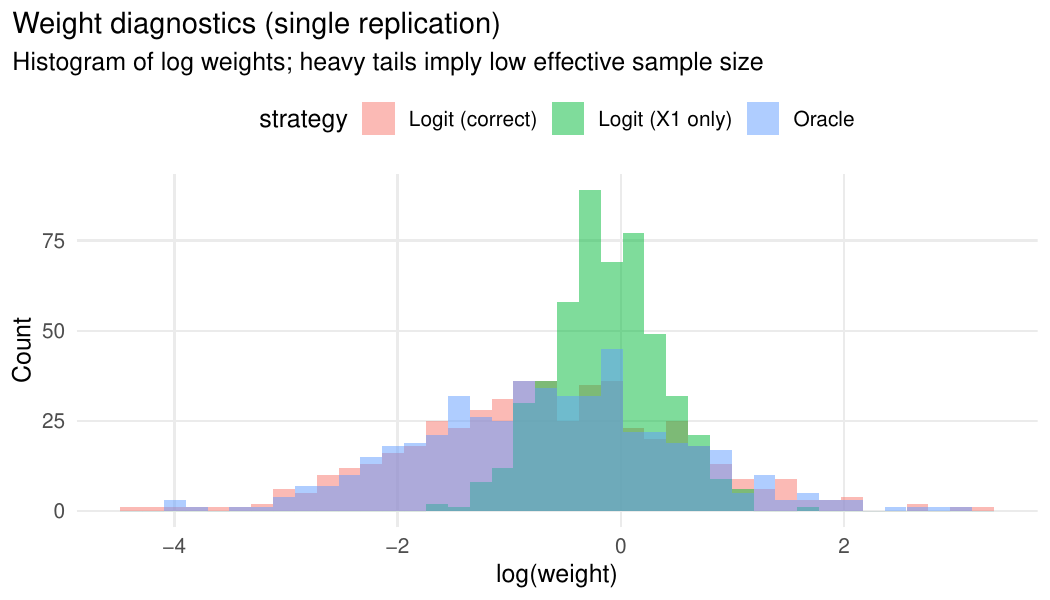}
\end{minipage}

\caption{Experiment 6 (DGP 1): sensitivity to weight estimation.
\emph{Top-left:} coverage by weighting strategy.
\emph{Top-right:} width by weighting strategy.
\emph{Bottom:} weight distribution diagnostic (log-scale), illustrating tail behavior across strategies.}
\label{fig:exp6-weights}
\end{figure}

\paragraph{Results.}
\Cref{fig:exp6-weights} shows that (in this moderate covariate-shift setting) bounds are
stable across weighting strategies: estimated weights closely match oracle performance,
and even mild misspecification does not drastically change width or coverage at these
$\Lambda_{\mathrm{out}}$ values. The weight histogram highlights how misspecification
changes tail behavior (maximum weight and effective sample size), which is useful for
diagnostics in applications.

\subsection{Experiment 6b: identification vs.\ estimation scaling (DGP 1)}
\label{sec:sim-exp6b}

\paragraph{Design.}
This experiment isolates an important failure mode of naive inference: even with very
large $n^r$, the transported point estimator can be \emph{precisely wrong} when outcome
transportability fails. We simulate DGP 1 with a large target covariate sample $n^o=5000$,
vary trial size $n^r\in\{200,500,1000,2000,5000\}$, and run $R=200$ replications per $n^r$.
We compare:
(i) the naive transported point estimate (equivalently $\Lambda_{\mathrm{out}}=1$),
(ii) a naive nonparametric bootstrap CI for the transported estimator (200 bootstrap resamples),
and (iii) our sharp bounds at $\Lambda_{\mathrm{out}}\in\{1.5,2.0\}$.

\begin{figure}[t]
\centering
\begin{minipage}{0.49\linewidth}
\centering
\includegraphics[width=\linewidth]{img/exp6_id_vs_est_coverage_vs_nr.pdf}
\end{minipage}\hfill
\begin{minipage}{0.49\linewidth}
\centering
\includegraphics[width=\linewidth]{img/exp6_id_vs_est_width_vs_nr.pdf}
\end{minipage}

\vspace{0.5em}

\begin{minipage}{0.70\linewidth}
\centering
\includegraphics[width=\linewidth]{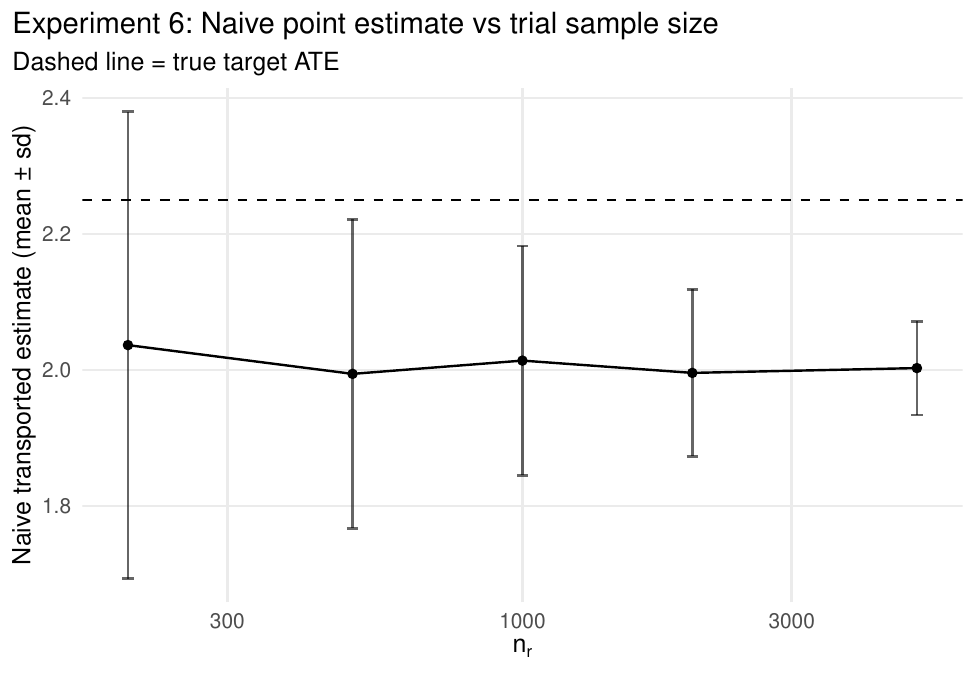}
\end{minipage}

\caption{Experiment 6b (DGP 1): identification vs.\ estimation. As $n^r$ grows,
the naive bootstrap CI becomes narrower yet less valid, while sharp bounds retain coverage
and stabilize in width. The bottom panel (optional) illustrates convergence of the naive
point estimate away from the true $\tau^o$.}
\label{fig:exp6-id-vs-est}
\end{figure}

\paragraph{Results.}
\Cref{fig:exp6-id-vs-est} shows a stark pattern: as $n^r$ increases, the naive bootstrap CI
shrinks rapidly but coverage \emph{worsens} (e.g., dropping to $\approx 0.07$ by $n^r=5000$),
demonstrating that the dominant limitation is \emph{identification}, not sample size.
In contrast, the sharp bounds maintain near-nominal coverage with widths that do not
collapse to zero (as expected under partial identification).

\subsection{Experiment 7: bounded-support DGP where the LR model holds literally}
\label{sec:sim-exp7}

A common technical concern is that for Gaussian outcomes the pointwise likelihood ratio
can be unbounded. To demonstrate that our method behaves as intended when the model holds
literally, we construct a bounded-support DGP (DGP 7) with truncated covariates and truncated
outcomes so that the conditional likelihood ratio is finite.

\paragraph{DGP 7 definition (bounded).}
We draw covariates from truncated normals:
\[
X^r \sim \text{TruncNormal}(0,I_p;\ [-B,B]^p),\qquad
X^o \sim \text{TruncNormal}(\delta\mathbf{1}_p,I_p;\ [-B,B]^p),
\]
with $B=3$ and $\delta=0.5$. The moderator is generated as before with $\gamma_r=0$ and
$\gamma_o=0.2$:
$U^s\mid X\sim \mathcal{N}(\gamma_s X_1,1)$.
Outcomes follow a truncated normal:
\[
Y(a)\sim \text{TruncNormal}\!\Bigl(\beta_0+\beta_x^\top X + a(\tau_0+\beta_u U),\ \sigma^2;\ [y_L,y_U]\Bigr),
\]
with $\beta_0=1$, $\beta_x=(0.5,0.3,0.2,0.1,0.1)^\top$, $\tau_0=1$, $\beta_u=0.25$,
$\sigma=1$, and $[y_L,y_U]=[-3,3]$. For this DGP we also compute a literal global
$\Lambda_{\mathrm{DGP}}$ by maximizing the conditional likelihood ratio over $(a,x,y)$
on the bounded support (reported in the corresponding table file).

\begin{figure}[t]
\centering
\begin{minipage}{0.49\linewidth}
\centering
\includegraphics[width=\linewidth]{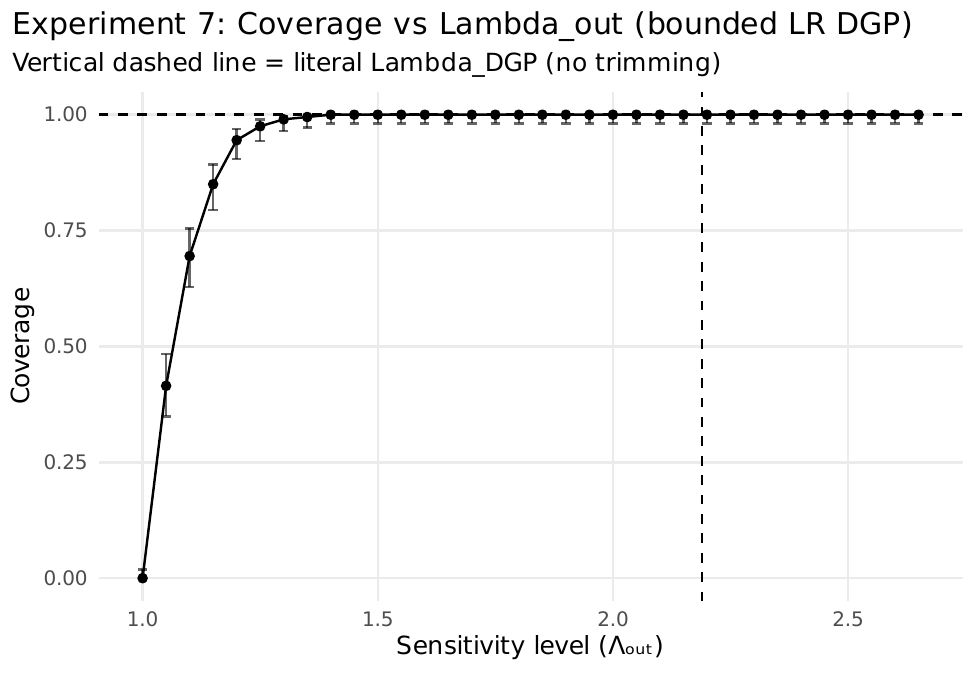}
\end{minipage}\hfill
\begin{minipage}{0.49\linewidth}
\centering
\includegraphics[width=\linewidth]{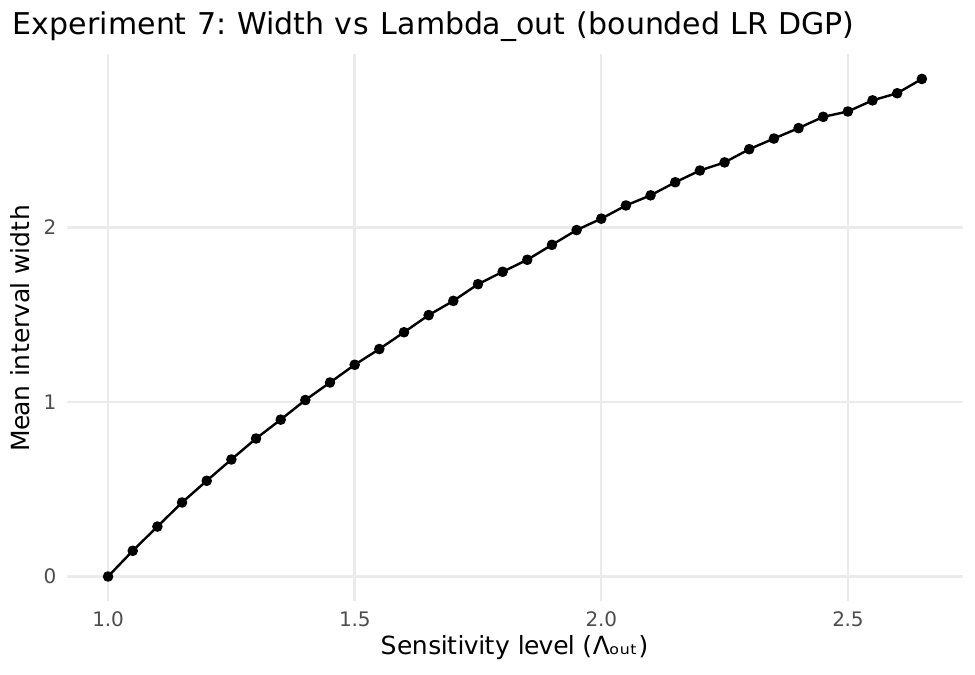}
\end{minipage}

\vspace{0.5em}

\begin{minipage}{0.70\linewidth}
\centering
\includegraphics[width=\linewidth]{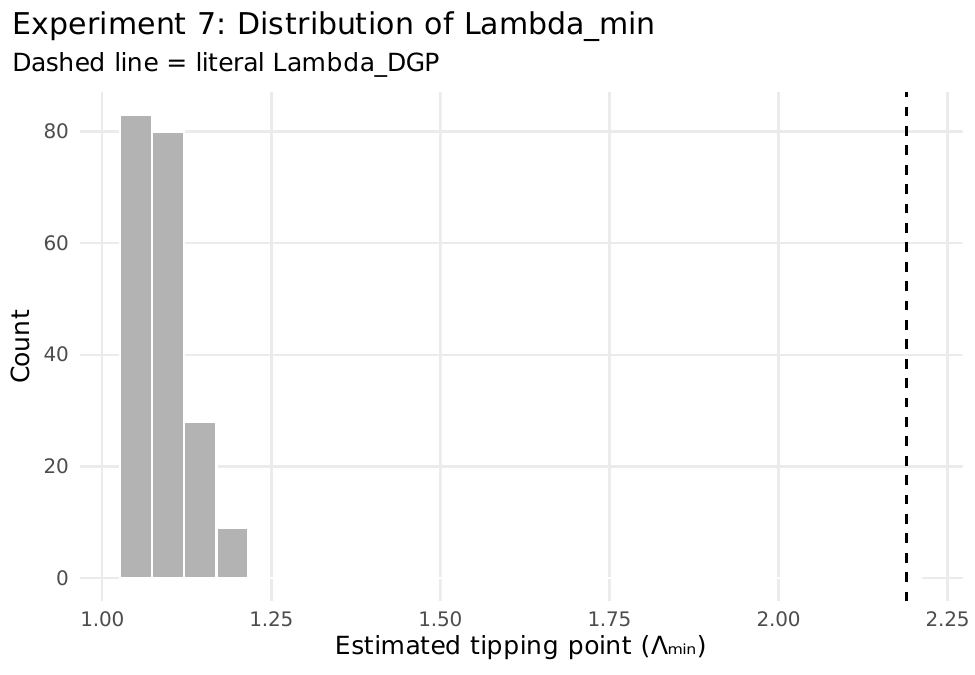}
\end{minipage}

\caption{Experiment 7 (bounded-support DGP): when outcomes and covariates have bounded
support, the pointwise LR model holds literally. Coverage increases with
$\Lambda_{\mathrm{out}}$ and the tipping-point distribution is concentrated.}
\label{fig:exp7-bounded}
\end{figure}

\paragraph{Design and results.}
We run an Exp-1-style sweep with $(n^r,n^o)=(500,1000)$ and $R=200$ replications over a
grid of $\Lambda_{\mathrm{out}}$. \Cref{fig:exp7-bounded} shows the same qualitative
pattern as in DGP 1, but here the LR-bounded model is well-defined without trimming:
coverage increases monotonically with $\Lambda_{\mathrm{out}}$, and widths grow smoothly.
The tipping points concentrate near $\Lambda_{\min}\approx 1.1$ in this particular setting.

\subsection{Experiment 8: ``bang--bang'' tilting visualization}
\label{sec:sim-exp8}

Finally, we visualize the structure predicted by \cref{lem:threshold}: the extremal
solutions that attain the sharp upper/lower bounds correspond to multipliers that
largely saturate at the LR constraints, producing a near-threshold (``bang--bang'')
pattern when ordered by outcome (or the relevant sufficient statistic).

\begin{figure}[t]
\centering
\includegraphics[width=0.9\linewidth]{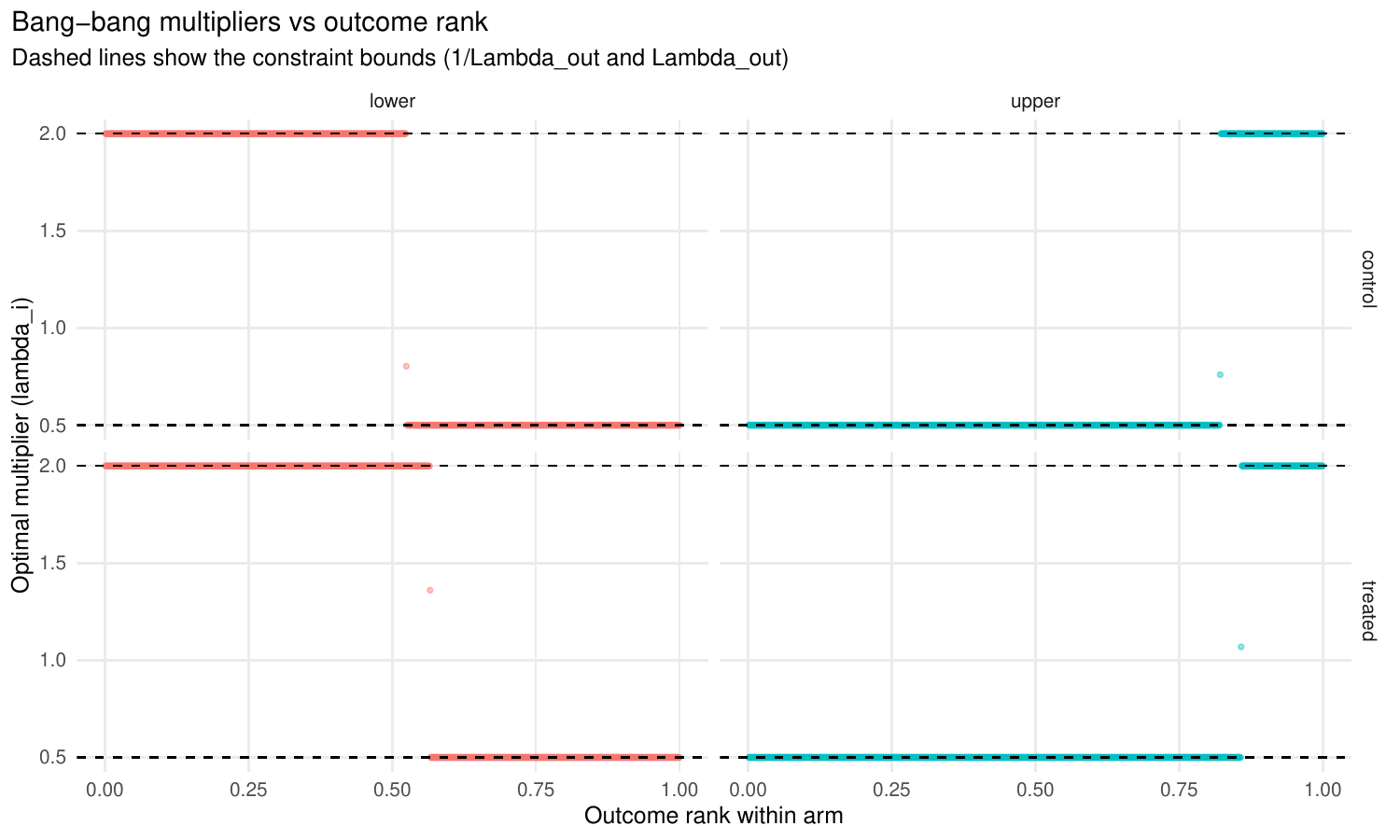}
\caption{Experiment 8: ``bang--bang'' structure of the bound-attaining tilts. The
estimated multipliers (density ratio adjustments) concentrate near the constraint
values and switch sharply around a threshold, consistent with \cref{lem:threshold}.}
\label{fig:exp8-bangbang}
\end{figure}

\subsection{Summary discussion (what the simulations show)}
\label{sec:sim-discussion}

Across DGPs and experimental designs, the simulations support four main conclusions.
First, the proposed procedure achieves empirical coverage once $\Lambda_{\mathrm{out}}$
is large enough to plausibly contain the true outcome shift, and it transitions smoothly
from the transported point estimate at $\Lambda_{\mathrm{out}}=1$ to wider partial-identification
intervals as $\Lambda_{\mathrm{out}}$ increases (Exp.~1, Fig.~\ref{fig:exp1-val}). Second,
the intervals are \emph{sharp} in the sense that their widths closely match large-$n$
oracle widths, implying little avoidable conservatism in finite samples (Exp.~1,
Tab.~\ref{tab:exp1-key}). Third, the method exposes an explicit coverage--informativeness
tradeoff and yields interpretable ``breakeven'' sensitivity levels under varying degrees
of moderator shift (Exp.~2, Tab.~\ref{tab:exp2-breakeven}). Fourth, the identification
aspect is central: naive transported estimators can become arbitrarily precise yet invalid
as $n^r$ grows, whereas sharp bounds maintain coverage and do not collapse (Exp.~6b,
Fig.~\ref{fig:exp6-id-vs-est}). Robustness experiments (Exp.~5) and diagnostics around
weight estimation (Exp.~6) further support practical applicability, while Exp.~7 and Exp.~8
address common technical and interpretability questions about LR-bounded models and the
structure of the optimizing tilts.

\end{document}